\documentclass[aps,prl,twocolumn,a4paper,groupedaddress,floatfix,10pt,longbibliography]{revtex4-1}

\usepackage{color,graphicx,amsmath,amssymb,amsfonts,caption,subcaption}
\usepackage{hyperref}

\newcommand{\Ng}{M}
\newcommand{\Ns}{N}
\newcommand{\ms}{n}
\newcommand{\Uf}{u}
\newcommand{\Rs}{R}
\newcommand{\Fs}{f}
\newcommand{\sm}{d}
\newcommand{\smaxO}{\Gamma^{1\text{-D}}}
\newcommand{\smaxZ}{\Gamma^{2\text{-D}}} 
\newcommand{\smaxD}{\Gamma^{3\text{-D}}}
\newcommand{\chiN}{d_{99}}
\newcommand{\chiNM}{\chi_{99}}
\newcommand{\mf}[1]{\boldsymbol{#1}}
\newcommand{\dims}{K}

\begin{document}  


\title{
A Quantum Inspired Approach to Exploit Turbulence Structures
} 

\author{Nikita Gourianov$^{1}$}
\email{nikgourianov@icloud.com}
\author{Michael Lubasch$^{2}$}
\author{Sergey Dolgov$^{3}$}
\author{Quincy Y.  van den Berg$^{1}$}
\author{Hessam Babaee$^{4}$}
\author{Peyman Givi$^{4}$}
\author{Martin Kiffner$^{5,1}$}
\author{Dieter Jaksch$^{1,5,6}$}
\email{dieter.jaksch@uni-hamburg.de}

\affiliation{$^{1}$Clarendon Laboratory, University of Oxford, Oxford, UK}
\affiliation{$^{2}$Cambridge Quantum Computing Limited, London, UK}
\affiliation{$^{3}$Department of Mathematical Sciences, University of Bath, Bath, UK}
\affiliation{$^{4}$Department of Mechanical Engineering and Materials Science, University of Pittsburgh, Pittsburgh, PA, USA}
\affiliation{$^{5}$Centre for Quantum Technologies, National University of 
Singapore, Singapore}
\affiliation{$^{6}$Insitut für Laserphysik, Universität Hamburg, Hamburg, Germany}

\begin{abstract}
Understanding turbulence is the key to our comprehension of many natural and technological flow processes. At the heart of this phenomenon lies its intricate multi-scale nature, describing the coupling between different-sized eddies in space and time. Here we analyse the structure of turbulent flows by quantifying correlations between different length scales using methods inspired from quantum many-body physics. We present results for interscale correlations of two paradigmatic flow examples, and use these insights along with tensor network theory to design a structure-resolving algorithm for simulating turbulent flows. With this algorithm, we find that the incompressible Navier-Stokes equations can be accurately solved even when reducing the number of parameters required to represent the velocity field by more than one order of magnitude   compared to direct numerical simulation. Our quantum-inspired approach provides a pathway towards conducting computational fluid dynamics on quantum computers.
\end{abstract}
\maketitle

Turbulence  can  phenomenologically be described by  
mutually interacting eddies stretching across 
an extremely broad range of length and time 
scales~\cite{monin:71,monin:75}. 
These spatial scales range from the largest size of the  energy 
containing eddies (known as the integral scale,  $\ell$), to the 
smallest ones, known as the Kolmogorov microscale, $\eta$. The 
separation between these scales is  $\ell / \eta 
\sim \text{Re}^{3/4}$ where $\text{Re}$ is the Reynolds 
number. A salient feature of turbulent flows is the scale-locality of the turbulent energy cascade~\cite{kolmogorov1941,Eyink2005,Cardesa2017,ChenShiyiETAL}, in the sense that eddies at a given length scale predominantly interact with other eddies of similar scale.

Since the pioneering work of Orszag and Patterson~\cite{OP72}, direct numerical simulation (DNS) of the 
Navier-Stokes equations has been widely regarded as the 
computational method with the highest fidelity in capturing the 
dynamics of turbulent flows. 
Virtually all classical DNS methods such as  spectral polynomial/Fourier, finite volume and finite element   are  \emph{scale-resolving}, where increasing the number of variables (e.g. grid points $\Ng$) resolves finer and finer scales. However, the wide separation of turbulent flow scales limits the range of Reynolds numbers that can be computationally considered. Straightforward estimates indicate that simulation of an incompressible flow inside a three-dimensional volume $\sim(10 \ell)^3$ with  $\text{Re} \sim 10^5$ would require  {\it decades} of CPU time on a 1 teraflop computer. 

For mitigating this huge numerical complexity, the 
importance of exploiting so-called coherent 
structures of turbulence~\cite{Hussain1983} 
has long been recognised. 
This eventually led to the rise of \emph{structure-resolving} methodologies (e.g., proper orthogonal decomposition)~\cite{HLB96} that extract and exploit correlated structures of the solution. They represent the flow field down to the Kolmogorov microscale through a superposition of modes, but with their number being much smaller than the total number of grid points in DNS. Up to now, the majority of the developed techniques have been used for \emph{diagnostic} purposes. Using reduced order models for \emph{predictive} purposes is hampered by difficulties in identifying suitable modes and 
remains an active area of research~\cite{AIAA17}. 

A similar challenge has been successfully 
tackled in a completely different area of 
physics.  Quantum many-body systems 
are described by elements of a vector space 
whose dimension grows \emph{exponentially} with the number of particles~\cite{Poulin2011,Orus2014}. This makes direct simulations quickly impossible with increasing system size.
Tensor network methods~\cite{Schollwock2011,verstraete:08} made a revolutionary advance in simulating quantum systems with local interactions by removing unrealised long-distance correlations, thus  enabling the 
simulation of physical systems that are otherwise 
intractable~\cite{clark:04,white:11}. 
The correlations of interest for quantum systems are known as 
quantum entanglement~\cite{horodecki:09}, and  
weakly entangled systems or those where  quantum correlations are concentrated at the boundaries between different parts of the system (i.e., they are structured to follow a so-called area law~\cite{Eisert2010}) can be  efficiently simulated with  
tensor network methods. 

Here we adapt these successful strategies for  treating 
quantum many-body systems towards exploiting the scale-locality of turbulence.  
This quantum inspired approach allows us to develop structure-resolving methods for both diagnostic and predictive purposes. We first introduce tools from tensor network theory to analyse different length scales of flow in a manner different from both the traditional wavenumber approach~\cite{onsager:49,weizsaecker:48,heisenberg:48,mathis2009} and more recent investigations in real space~\cite{Cardesa2017}. Then, we apply these diagnostic tools to study DNS solutions of two paradigmatic flow configurations: a 2-D temporally developing jet (TDJ)~\cite{Givi89} and the 3-D Taylor-Green vortex (TGV)~\cite{Brachet1983}. Our study reveals that the scale-locality of turbulence restricts the amount of correlation present between different length scales. This motivates us to encode turbulence in a simple tensor network called matrix product state (MPS). The connectivity of the MPS network is well adapted to describing scale-local flows and hence exploiting the structures of turbulence. We design an algorithm for simulating the incompressible Navier-Stokes equations (INSE) in the compressed MPS format. This algorithm remains accurate  even when reducing the number of variables parametrising the solution (NVPS)  by more than one order of magnitude. The conceptual similarity between the tensor network algorithm presented here and those used in quantum physics opens the possibility of conducting computational fluid dynamics on a quantum computer.
\section{Results \label{results}}
\begin{figure*}[t!]\begin{center}
\includegraphics[width=.98\linewidth]{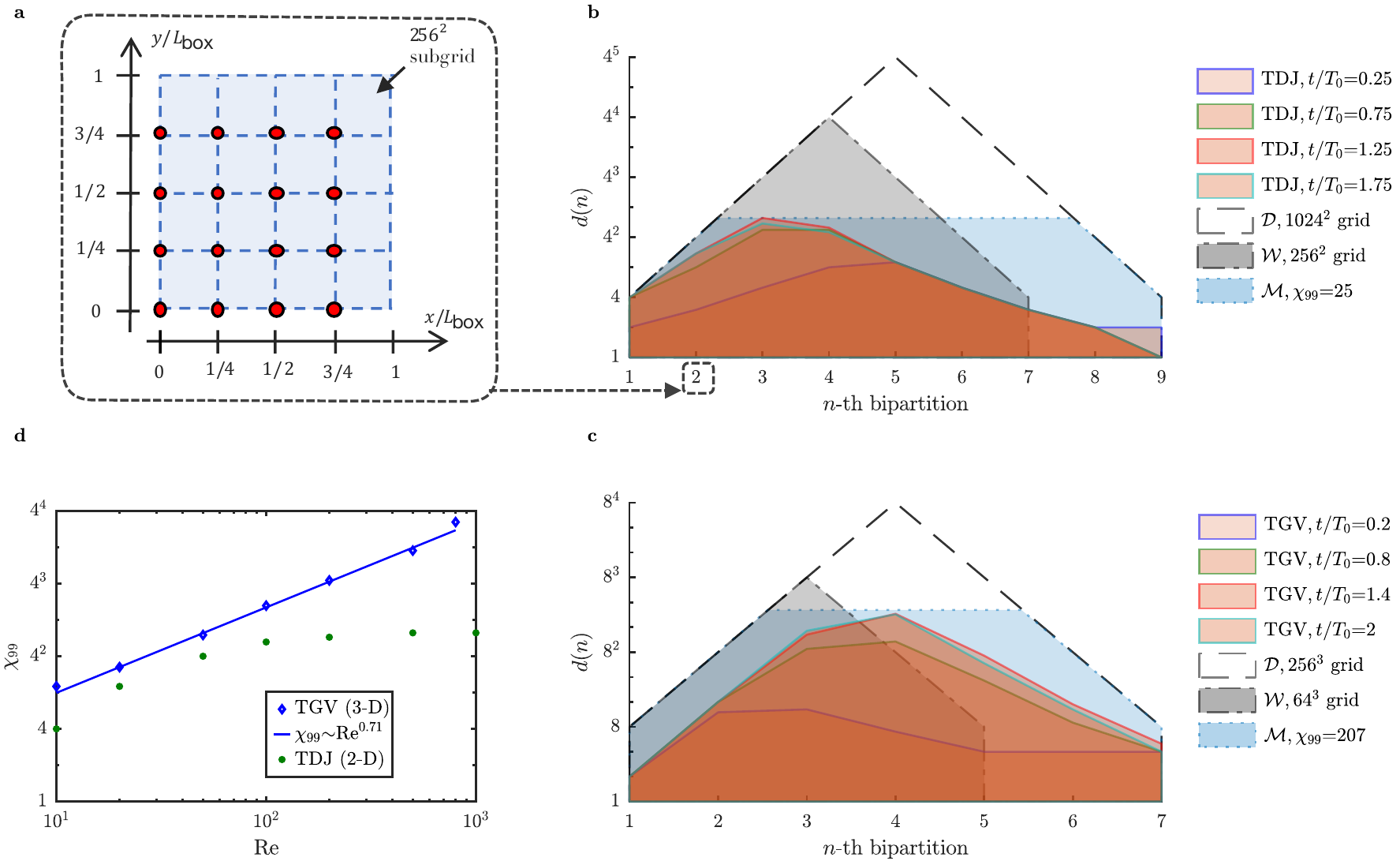}
\caption{\label{fig:CompSpaces} \textbf{Interscale correlations of turbulent fluid flows}. (\textbf{a}) and (\textbf{b}) correspond to a square with edge length $L_{\text{box}}$ on a $2^{10}\times2^{10}$  grid. (\textbf{a}) illustrates the subgrid structure when decomposing a  function $\Uf_i$ according to Eq.~(\ref{mpsScale}) for $\ms=2$. Red dots are the $2^{2}\times2^{2}$ grid points $\mf{X}_k$ of the coarse grid. Each blue square attached to the $\mf{X}_k$ indicates the quadratic subgrid with the $2^{8}\times2^{8}$ grid points $\mf{x}_k$ of the fine grid. (\textbf{b}) shows the Schmidt numbers $\chiN(n,t)$ on a logarithmic scale such that the decomposition in Eq.~(\ref{mpsScale}) results in a $99\%$ accurate representation of DNS solutions to the INSE at four different times [see Fig.~\ref{fig:sims2D}]. The domain $\mathcal{D}$ indicated by the black dashed line corresponds to  DNS. The grey shaded area $\mathcal{W}$ is for solutions on  a $2^8\times2^8$ grid. The blue shaded area $\mathcal{M}$ is for $\sm(n)\le25$ in Eq.~(\ref{mpsScale}). (\textbf{c}) Same as in (\textbf{b}) but for the 3-D simulations shown in Fig.~\ref{fig:sims3D}. In (\textbf{b}) and (\textbf{c}), $T_0$ is the characteristic time scale on which the quickest particles in the initial flow fields can traverse the box (see \emph{Set-up of numerical experiments} section in \emph{Methods}). (\textbf{d}) Scaling of  $\chiNM=\max \chiN(\ms,t)$ with the Reynolds number for the 2-D and 3-D systems in (\textbf{b}) and (\textbf{c}), respectively.}
\end{center}
\end{figure*}

\textbf{Quantifying interscale correlations.} Throughout this work we follow the standard approach 
in computational fluid dynamics and discretise 
the computational domain. 
Each spatial dimension is discretised by 
$2^{\Ns}$ grid points, where $\Ns$ is a positive integer. 
In this way, the velocity field
\begin{align}
\mf{V}(t,\mf{r}_q) = 
\sum\limits_{i=1}^{\dims} \Uf_i(t,\mf{r}_q) \hat{\mf{e}}_i 
\end{align}
and its  Cartesian components $\Uf_i$ 
 are discrete functions of the grid points $\mf{r}_q$, 
 where $\dims$ is the spatial dimension 
 and $\hat{\mf{e}}_i$ are Cartesian unit vectors. 
We measure the interscale correlations by using the Schmidt (singular value) decomposition to systematically divide 
the computational grid into sub-grids, as illustrated in Fig.~\ref{fig:CompSpaces}a for $K=2$.
We decompose 
(for details, see the \emph{Schmidt decomposition} section in \emph{Methods}) each component $\Uf_i$ on 
this $2^{\Ns}\times2^{\Ns}$ grid   
into functions  $R$ and $f$  on 
a coarse and a fine subgrid, respectively, 
\begin{align}
 \Uf_i(t,\mf{r}_q)=
 \sum\limits_{\alpha=1}^{\sm(\ms)}\lambda_{\alpha}(t) \Rs_{\alpha}(t,\mf{X}_k)\Fs_{\alpha}(t,\mf{x}_l), \quad \mf{r}_q=\mf{X}_k+\mf{x}_l.
 \label{mpsScale}
\end{align}
Positions $\mf{X}_k$ correspond to a quadratic grid with $2^{\ms}\times2^{\ms}$  
points (coarse grid), and $\mf{x}_l$ correspond to 
a fine sub-grid with $2^{\Ns-\ms}\times2^{\Ns-\ms}$ grid points. 
The functions $\Rs_{\alpha}$ and $\Fs_{\alpha}$ obey the 
orthonormality condition 
\begin{align}
\sum_k \Rs_{\alpha} (t,\mf{X}_k) \Rs_{\beta} (t,\mf{X}_k) = \sum_l 
\Fs_{\alpha}(t,\mf{x}_l) \Fs_{\beta}(t,\mf{x}_l) = \delta_{\alpha \beta}\,,
\end{align}
where $\delta_{\alpha \beta}$ is the Kronecker delta. 
 The parameter $\ms=1,\ldots,\Ns-1$ labels the 
 possible bipartitions 
 of the square lattice in coarse and fine grids [see 
 Fig.~\ref{fig:CompSpaces}a for $\Ns=10$ and $\ms=2$]. 
 The Schmidt number $\sm(\ms)$ denotes 
the number of retained terms in the summation in 
Eq.~(\ref{mpsScale}), 
and each product $\Rs_{\alpha}\Fs_{\alpha}$ 
is weighted by a Schmidt coefficient $\lambda_{\alpha}\ge0$.  
These coefficients appear in descending order 
$\lambda_1\geq \lambda_2... \geq \lambda_{\sm(\ms)}$, 
thus varying $\sm(\ms)$ will only 
add or remove the least important of the 
orthonormal basis functions. 
Here we take $\sm(\ms)$ as a  quantitative measure 
for the interscale correlations of turbulent flows at a given 
bipartition $\ms$ of the lattice: 
$\sm(\ms)=1$ corresponds to an uncorrelated product state,
and with increasing  $\sm(\ms)$ the flow 
becomes more strongly correlated  between the coarse and the fine grid.
Note that while the $\sm(\ms)=1$ product 
state exhibits no interscale correlations, 
it is still highly correlated in space because the fine grid 
dependence is repeated.

Truncating the Schmidt decomposition in 
Eq.~(\ref{mpsScale}) approximates 
$u_i$ in an orthonormal time-dependent basis 
that evolves with the fluid flow to optimally capture spatially 
correlated structures. 
This is in 
contrast to classical scientific computing techniques (implemented 
through e.g. finite difference or spectral methods) where the bases 
are structure-agnostic, i.e., 
they are chosen a priori and disregard any 
structure in the solution. 

\begin{figure*}[t!]\begin{center}\includegraphics[width=.98\linewidth]{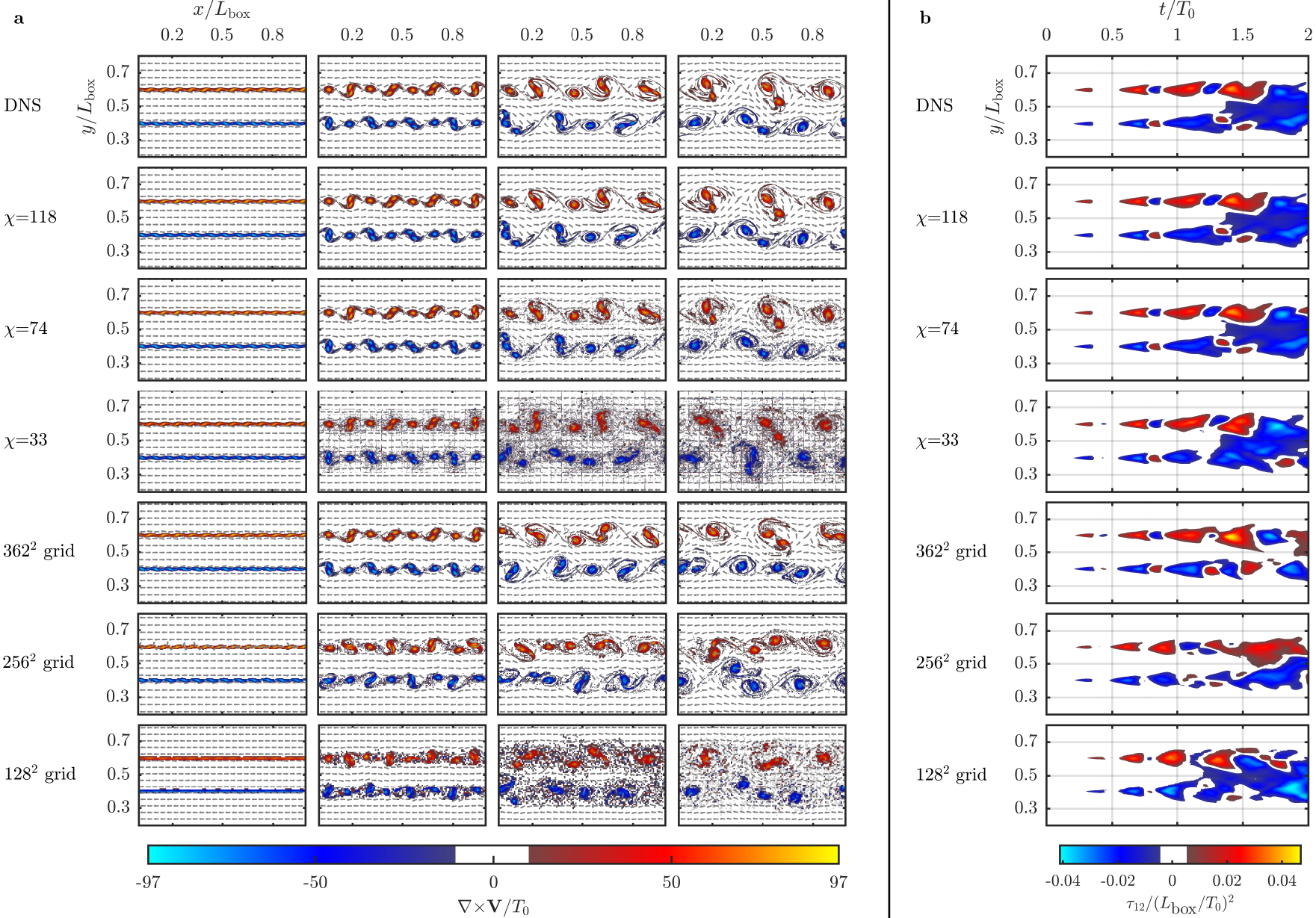}
\caption{\label{fig:sims2D} \textbf{$\mathbf{2}$-D Temporally developing jet.} Dynamical simulation of the INSE in 2-D for a planar jet streaming along $x$ with $\text{Re}=1000$, as defined in the \emph{Set-up of numerical experiments} section in \emph{Methods}. (\textbf{a}) shows  snapshots of the vorticity and velocity fields taken at $t/T_0 = 0.25, 0.75, 1.25, 1.75$ (left to right). Red corresponds to positive vorticity (counter-clockwise flow) and blue to negative (clockwise). Top row corresponds to DNS results on a quadratic $2^{10}\times2^{10}$ grid (cf. Fig.~1a). Rows 2-4 are MPS results with different maximal bond dimensions $\chi$. Bottom three rows are for URDNS on quadratic grids as indicated. (\textbf{b}) Reynolds stress $\tau_{12}$ [see Eq.~\eqref{stress}] between the streamwise  and cross-stream directions as a function of time and $y$ coordinate. Red (blue) corresponds to positive (negative) stress.}
\end{center}
\end{figure*}

We first apply the decomposition in Eq.~(\ref{mpsScale}) 
to  DNS solutions of the INSE 
[see Eq.~(\ref{eq:INSE0})] for the TDJ shown in the top row of 
Fig.~\ref{fig:sims2D}a. 
The TDJ comprises a central jet flow along the $x$-direction, 
and  Kelvin-Helmholtz instabilities in the boundary layer of 
the jet eventually cause it to collapse (see Eqs.~\eqref{eq:TDJ0} through~\eqref{eq:ensemAvg} in \emph{Methods} for initial flow conditions). 
We decompose each velocity component according to 
Eq.~(\ref{mpsScale}), which is an exact representation 
if 
\mbox{$\sm(\ms)=\smaxZ(\ms)$} with (for details, see Supplementary Sec.~2)
\begin{align}
 \smaxZ(\ms)=\min(4^\ms,4^{\Ns -\ms})\,.
 \label{gamma2d}
\end{align}
Fig.~\ref{fig:CompSpaces}b shows  the Schmidt numbers $\chiN(\ms,t)$ 
such that Eq.~(\ref{mpsScale}) represents the DNS solutions 
for the velocity fields with $99\%$ accuracy in the L2 norm 
(more details on the Schmidt coefficients
can be found in Supplementary Sec.~1). 
We find that  $\chiN(\ms,t)$ are well below their maximal values 
$\smaxZ(\ms)$ for $\ms>1$. 
More specifically, we define  $\chiNM=\max \chiN(\ms,t)$ as the 
maximal value of   $\chiN$ for all $\ms$ and time 
steps. We obtain $\chiNM=25$, and  the 
interscale correlations captured by  Eq.~(\ref{mpsScale}) with 
$\sm(\ms)=\min\left(\smaxZ(\ms),25\right)$ 
are shown by the blue-shaded area $\mathcal{M}$ in 
Fig.~\ref{fig:CompSpaces}b. 
$d_{99}(\ms,t)$ is entirely contained within this blue area. 
Note that the Schmidt numbers are shown on a logarithmic scale in 
Fig.~\ref{fig:CompSpaces}b, and thus the  area $\mathcal{M}$ 
is much smaller than the area $\mathcal{D}$ corresponding to DNS. 

\begin{figure}[t!]
\begin{center}
\includegraphics[width=.85\linewidth]{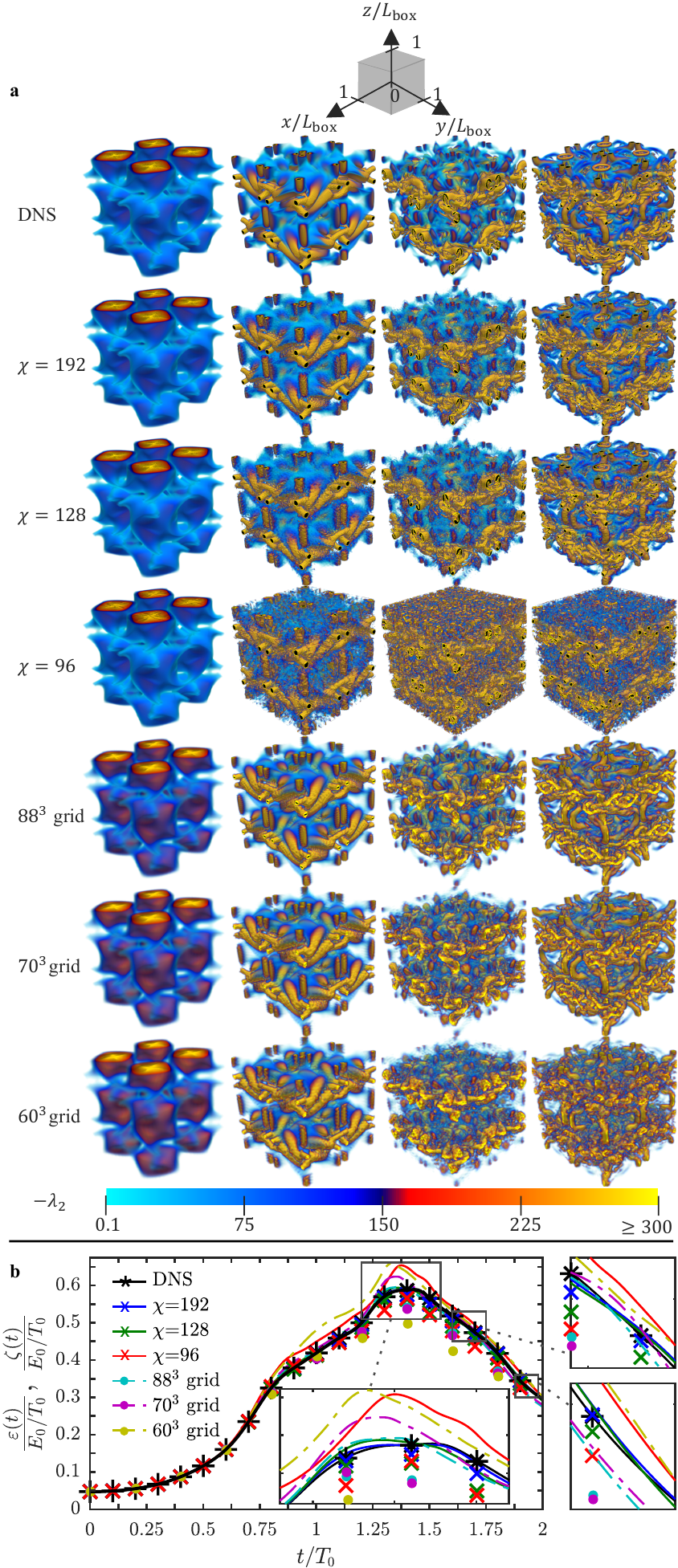}
\caption{\label{fig:sims3D}
\textbf{$\mathbf{3}$-D Taylor-Green vortex.} Dynamical simulation of the INSE in 3-D for the Taylor-Green vortex and $\text{Re}= 800$, as defined in the \emph{Set-up of numerical experiments} section in \emph{Methods}. (\textbf{a}) Vortical structures rendered using the standard $\lambda_2$ method \cite{Jeong1995} are shown at times $t/T_0=0.2,0.8,1.4,2$ (left to right). Top row is for DNS  on a $2^8\times2^8\times2^8$ grid. Rows 2-4 are for MPS simulations with different  $\chi$,  and bottom three rows are for URDNS  on cubic grids as indicated. (\textbf{b}) shows the enstrophy $\zeta(t)$ (asterisks/crosses/circles) and the energy dissipation $\epsilon(t)$  (lines) as a function of time, with $E_0$ being the total energy at $t=0$.}
\end{center}
\end{figure}

We obtain   qualitatively similar results 
for the DNS solutions to the TGV in 3-D, where 
vortex stretching causes a single, large, ordered fluctuation to 
collapse into a turbulent flurry of small scale 
structures (see top
row in Fig.~\ref{fig:sims3D}a for visualisation, and \eqref{eq:TGV0} in \emph{Methods} for initial flow conditions). 
In three spatial dimensions, the representation 
in Eq.~(\ref{mpsScale}) is exact  if $\sm(\ms)$ 
equals  (for details, see Supplementary Sec.~2)
\begin{align}
\smaxD(\ms) = \min(8^\ms,8^{\Ns-\ms})\,. 
\label{gamma3d}
\end{align}
The Schmidt numbers $\chiN(\ms,t)$ resulting 
in a $99\%$ accurate representation of the DNS solutions  
are shown in Fig.~\ref{fig:CompSpaces}c. 
We find $\chiNM=207$ such that all values of 
$\chiN$ in Fig.~\ref{fig:CompSpaces}c 
are contained within the blue-shaded area $\mathcal{M}$
corresponding to $\sm(\ms)=\min\left(\smaxD(\ms),207\right)$ in Eq.~(\ref{mpsScale}). Since $\chiNM$ is much smaller 
than the upper vertex of the area $\mathcal{D}$ at 
$\smaxD(4)=2^{12}$, 
the interscale correlations of DNS solutions 
are far from being saturated 
(more details on the Schmidt coefficients
can be found in Supplementary Sec.~1). 

Next we investigate how the maximal Schmidt number $\chiNM$ 
scales with the Reynolds number Re, and the results are  shown 
in Fig.~\ref{fig:CompSpaces}d. 
We find that $\chiNM$ saturates in the 2-D case for 
$\text{Re}\gtrsim200$. This 
suggests that interscale correlations of 2-D flows are bounded in 
analogy to quantum correlations in gapped 1-D quantum 
systems with local interactions~\cite{Eisert2010}.  
In the 3-D case,  $\chiNM$ increases according to a power law.  
The NVPS for $\sm(\ms)=\min\left(\smaxD(\ms),\chiNM\right)$ 
scales as $\sim\chiNM^2 \log \Ng$ (see Supplementary Sec.~2).  
Kolmogorov’s theory~\cite{kolmogorov1941}  states that the number of grid points $\Ng=8^\Ns$ must scale with the Reynolds number like 
$\Ng \sim (\ell/\eta)^3 \sim \text{Re}^{9/4}$ to resolve all spatial scales.
This implies that the NVPS of $\mathcal{M}$
only scales as $\sim\text{Re}^{1.42} \log \text{Re}$, 
which is a substantially slower increase with Re compared to the NVPS of DNS, $\sim\Ng  \sim \text{Re}^{9/4}$.

\textbf{Tensor network algorithm.}
The previous results demonstrate that it is 
beneficial to find a representation 
of flow fields where limiting the amount of 
interscale correlations directly translates into a  
reduction of the NVPS. 
This can be achieved by expressing each velocity component 
in a compressed tensor network format known as matrix 
product state (MPS) 
or tensor train 
decomposition~\cite{Schollwock2011,verstraete:08,oseledets:11,Holtz2012}.
Our MPS encoding of function values is chosen such that it  
is consistent with the decomposition in Eq.~(\ref{mpsScale}) 
(see Supplementary Sec.~2).   It comprises products of 
$\Ns$ matrices $A^{\omega_{\ms}}$ with dimension $\sm(\ms-1)\times\sm(\ms)$,  where $2^\Ns$ is the 
number of grid points in each spatial 
direction, $\ms=1,\ldots,\Ns$ and $\sm(0)=\sm(\Ns)=1$~\cite{Lubasch2018,Ripoll2021}. 
The matrix $A^{\omega_{\ms}}$ is associated with a length scale 
$L_{\text{box}}/2^{\ms}$, and its dimension $\sm(\ms)$ controls the maximum 
amount of 
correlations allowed  between neighbouring scales. 
The nearest neighbour correlations are mediated directly 
by each matrix product, while correlations between further distant length scales can only be captured indirectly by traversing 
several matrix products. 
These properties make MPS well suited for the description of 
scale-local turbulent flows where correlations between vastly different 
length scales are expected to be small.

Here we consider MPSs of bond dimension $\chi$ where 
we set $\sm(\ms)=\min\left(\smaxZ(\ms),\chi\right)$  
in 2-D and $\sm(\ms)=\min\left(\smaxD(\ms),\chi\right)$ in 
3-D. 
The bond dimension $\chi$ controls the level of compression 
in the MPS format.
For example, the interscale correlations captured by 
an MPS with bond dimension $\chi=25$ ($\chi=207$) are represented 
by the blue-shaded area in Fig.~\ref{fig:CompSpaces}b 
(Fig.~\ref{fig:CompSpaces}c).
If $\chi$ is kept constant as $\Ns$ 
increases, the number of MPS parameters scales logarithmically 
with the total number of grid points, resulting in an 
exponential reduction of the NVPS compared to DNS.
However, we emphasise that this reduction does not truncate the range of length scales
covered by the MPS ansatz, it only limits the amount of 
interscale correlations. 
In order to fully utilise this dimensionality reduction for 
numerical simulations on large grids, 
we devise an algorithm for solving the  INSE 
without  leaving the compressed MPS manifold $\mathcal{M}$ 
(see the \emph{Matrix product state algorithm} section in \emph{Methods}). 
 We use 
a second-order Runge-Kutta time-stepping scheme and 
discretise  spatial derivatives in the same 
way as the DNS solver, which utilises  an $8$th-order 
finite difference stencil.

\textbf{Validation of the tensor network algorithm.} We now investigate how well the dynamics of  turbulent flow are  captured inside the 
MPS manifold $\mathcal{M}$ by 
comparing our algorithm 
against DNS for different compressions. 
Reducing the bond dimension $\chi$ reduces the NVPS.
The analogue of reducing the bond dimension in traditional DNS is to perform underresolved DNS (URDNS) where the simulation is carried out on a coarse grid not covering all relevant length scales.
URDNS can be considered as the most basic form  
of large eddy simulations~\cite{Sagaut2006,Pope2004,Zhiyin2015}, where no explicit models are employed to 
account for the disregarded subgrid scales 
(see the \emph{Direct numerical simulation algorithm} section in \emph{Methods}). 
For a fair comparison between MPS and URDNS,  we choose for 
every bond dimension $\chi$ a corresponding URDNS grid 
 such that the NVPS 
is approximately equal for both methods. 

The results for the TDJ and for the different solvers are shown in Fig.~\ref{fig:sims2D}. The top row in Fig.~\ref{fig:sims2D}a corresponds to DNS and illustrates how the background perturbations in the shear-layer are amplified ($t/T_0=0.25$) until the layer rolls up into vortices which in turn pair-up and merge into progressively larger vortices ($t/T_0=0.75,1.25$) until $t/T_0=1.75$, when pairing is terminated. The stress exerted upon the mean flow by turbulent fluctuations is given by the Reynolds stress tensor, one of whose components is plotted in Fig.~\ref{fig:sims2D}b. These results are in  accord with the Boussinesq approximation which indicates the Reynolds stresses are of the opposite sign of the mean streamwise velocity gradients along the cross-stream direction. The exception is when the vortex pairing is terminated, and the growth of the shear-layer is temporarily paused. Correctly resolving $\tau_{12}$ is important for the physical validity of the simulation. 

We now evaluate the accuracy of the MPS and URDNS simulations. Rows 2-4 in Fig.~\ref{fig:sims2D} show MPS results for $\chi = 33,74$ and 118, corresponding to compression ratios of approximately $1:64$, $1:16$ and $1:8$ compared to DNS, respectively. These results show that the large-scale dynamics of the jet are correctly captured by all the MPS simulations, with $\chi=74\ \text{and}\ 118$ practically indistinguishable from DNS. The bottom three rows in Fig.~\ref{fig:sims2D} show the URDNS results for different grid sizes. We find that  the $362^2$ grid (corresponding to $\chi=118$) is only accurate until $t/T_0\approx 1.2$ while the lower-resolution URDNS already fails at $t/T_0\approx 0.75$. This is observed in both the instantaneous vorticity dynamics results and the Reynolds-stresses. The MPS algorithm achieves a much higher accuracy than URDNS for the same NVPS (see Table~\ref{tab} for quantitative results). This result can also be understood in terms of the interscale correlations shown in Fig.~\ref{fig:CompSpaces}b, where the domain corresponding to URDNS on the $256^2$ grid is shown by the grey-shaded area $\mathcal{W}$. A significant amount of correlations present in the DNS solutions are outside of $\mathcal{W}$, which is consistent with our finding that URDNS cannot accurately represent the DNS solutions.


Next we discuss the corresponding results for the dynamics of 
the TGV.
The top row in  Fig.~\ref{fig:sims3D}a corresponds  to DNS  
and illustrates how the original vortex 
collapses ($t/T_0= 0.2$) into turbulent worm-like structures 
($t/T_0=0.8$) which become progressively more turbulent ($t/T_0=1.4$) 
until viscosity eventually dissipates these vortical structures ($t/T_0=2$).
Rows 2-4 and 5-7 in Fig.~\ref{fig:sims3D}a correspond to the 
results of our MPS algorithm and to  URDNS, respectively. 
The bond dimension and grid sizes have been chosen such 
that the compression ratios compared to DNS are approximately
$1:25$ (rows 2 and 5), $1:49$ (rows 3 and 6) and $1:78$ (rows 4 and 7). 
While  MPS produces a solution comparable to DNS for a compression 
ratio of $1:49$ ($\chi=128$), the corresponding 
URDNS results clearly deviate from DNS. Discrepancies 
between URDNS and DNS are even visible for the largest 
URDNS grid (compression $1:25$). 

A more quantitative analysis of the performance of  MPS vs. URDNS 
is shown in Fig.~\ref{fig:sims3D}b. 
In the non-DNS simulations, a 
portion of the energy is erroneously lost to numerical 
diffusion. The amount of numerical diffusion 
can be  measured by comparing  the physical 
global dissipation (enstrophy) to the  global 
kinetic energy dissipation [see Eqs.~\eqref{eq:dedt} and~\eqref{eq:enst}]. 
The MPS predictions at $\chi\ge128$ are consistent 
with both the DNS results here and in 
previous work~\cite{Brachet1983}.
We find that the MPS simulations with 
$\chi=128\ \text{and}\ 192$ dissipate the 
energy more accurately than any of the URDNS results, especially for $t/T_0 \geq 1.4$, see Table~\ref{tab}.
As in the 2-D case, this outcome is in line with 
the interscale correlations shown in Fig.~\ref{fig:CompSpaces}c, 
where the domain corresponding to URDNS on the $64^3$ grid 
is shown by the grey-shaded area $\mathcal{W}$. 
The bipartitions at $n=5,6,7$ are associated with 
comparatively large Schmidt numbers for $t/T_0\geq0.8$, and 
hence these interscale correlations cannot be captured 
by URDNS. 

\section{Discussion \label{discussion}}
The structure-resolving properties of MPSs can lead to a reduced computational cost. The computational complexity of our MPS algorithm is $\sim \chi^4 \log \Ng$, as explained in Supplementary Sec.~4. Resolving down to the Kolmogorov microscale $\eta$ requires $\Ng \sim \left(\ell/\eta\right)^K \sim \textnormal{Re}^{3K/4}$ grid points. Assuming $\chi \sim\textnormal{Re}^{\gamma}$, the overall scaling of the MPS algorithm becomes $\sim\textnormal{Re}^{4\gamma} \log\textnormal{Re}$. Comparing this to the scaling of DNS, which is $\sim M \log M  \sim\textnormal{Re}^{3 K/4} \log \textnormal{Re}$ (see the \emph{Direct numerical simulation algorithm} section in \emph{Methods}), we see that the MPS algorithm outperforms DNS when $\gamma < 3K/16$. For our 2-D example, Fig.~1d suggests $\gamma \approx 0$ which leads to an exponential speedup of the MPS algorithm over DNS for sufficiently large Re. It would be interesting to investigate whether this saturation of the Schmidt number with Re is a unique case for just the TDJ flow, or if it is a more general property of 2-D turbulence. If this is indeed general, it would have significant practical consequences for e.g. the simulation of atmospheric flows. For the TGV flow $\gamma \approx 0.71$, which is larger than $3K/16$. However, we note that numerical methods for manipulating high-dimensional tensors are an active field of research [Supplementary Sec.~4D], potentially enabling an improved scaling of MPS algorithms with $\chi$ in the future. We also remark that MPS can be exponentially faster than DNS at simulating shock waves, as illustrated in Supplementary Sec.~3 through analytical studies of the 1-D Burgers’ equation.


Our initial choice of MPS networks was motivated by the scale-locality of turbulent flows. 
However, MPS might be numerically inefficient when correlations between distant scales are relevant. One 
then needs a very large  bond dimension $\chi$ 
to maintain an 
accurate description of the flow. 
Other tensor network geometries like Tree 
Tensor Networks (TTNs) or 
multi-scale  
entanglement renormalisation ansatz (MERA) and its derivatives~\cite{Vidal2007,Evenbly2014} 
might then  be worthwhile considering.
These network geometries 
(see Supplementary Sec.~6) have direct 
bonds between further distant length 
scales and might require smaller bond dimensions. 
However, TTNs  maintain numerical
efficiency by abandoning direct bonds 
between neighbouring length scales which are important for exploiting scale-locality. MERA is numerically challenging because of loops in the network.

The utility of tensor networks in fluid dynamics goes beyond the 
INSE. Future avenues of investigation for MPS include compressible 
flows, in which the Mach number is an important parameter, and 
transport of scalar quantities under both passive and chemically 
reactive conditions where the effects of Prandtl, Peclet and 
Damkohler numbers~\cite{LW94} must be taken into account. It would be 
interesting to examine how these parameters affect the fidelity of 
low $\chi$ MPS simulations. Furthermore, as tensor network methods 
are naturally suited to tackle high-dimensional problems, their 
applicability to the transported probability density function (PDF) 
of turbulent reactive flows~\cite{Pope85} should be considered. In 
these flows, in addition to temporal and spatial variations, the PDF 
is a function of the three-dimensional velocity field and all of the 
pertinent scalar variables (energy, pressure and species mass 
fractions)~\cite{NNGLP17PRF}. With just $10$ species (a very simple 
chemical kinetics model), the unsteady PDF must be resolved in a $17$-
dimensional space. High-fidelity modelling and simulation of such 
complex flows can potentially be enabled through a well-chosen tensor 
network ansatz.

The close connection of our tensor network-based approach to 
quantum physics points towards the prospect of solving 
the Navier-Stokes equations on a quantum computer. Recently, several algorithms 
for solving nonlinear partial differential equations on quantum 
computers have been proposed~\cite{lubasch:20,lloyd:20,Liu2021}. In particular,
the work in~\cite{lubasch:20} introduces tensor networks as a programming 
paradigm for quantum computers, which makes our approach especially 
well-suited for quantum hardware implementations (see Supplementary Sec.~5 for details).  
Replacing classical floating point operations by quantum gates reduces the  scaling with bond dimension to $\sim\chi^2$ (see Supplementary Sec.~5). In addition, potentially exponential speedups are possible by choosing an optimised quantum network  that 
goes beyond the MPS ansatz for encoding the solution~\cite{Pollman2020, Pollman2021, lubasch:20}. In this way, our work holds the promise of enabling large-scale computational fluid dynamics calculations that are well beyond the scope of current approaches.

\section{Methods}
\textbf{Schmidt decomposition.}  
We consider a 1-D system and scale all lengths 
with its spatial dimension $L_{\text{box}}$.
We discretise the spatial domain $[0,1]$ of the velocity 
$\Uf$ with $\Ns$ bits into $2^\Ns$ grid points 
$r_q = q/2^{\Ns}$ with  $q = 0,1,\ldots,2^\Ns{-}1$. 
Next, we introduce $\ms = 1,\ldots,\Ns{-}1$ bipartitions of this 
grid into coarse and fine sub-grids. For a given $\ms$, 
the coarse sub-grid comprises the points $X_k = k/2^{\ms}$ with 
$k = 0,\ldots,2^\ms{-}1$. The spacing between neighbouring 
points is thus $2^{-\ms}$ and this defines the coarse length 
scale. To each coarse grid point $X_k$ is attached a fine 
sub-grid with points $x_l = l/2^{\Ns}$ with  
$l = 0,1,\ldots,2^{(\Ns-\ms)}{-}1$, and adjacent points 
are separated by the fine length scale $2^{-\Ns}$. 
In this way, any point $r_q$ of the 1-D grid can be written 
as $r_q = X_k + x_l$. Finally, we arrange the function values 
\mbox{$u(r_q) = u(X_k + x_l)$}  into a 
$2^n\times 2^{\Ns-\ms}$ matrix 
where the rows  and columns correspond to increments along 
the coarse and fine grids, respectively. 
Performing a singular value decomposition (SVD) on 
this matrix~\cite{Schollwock2011, Orus2014} gives the desired 
Schmidt decomposition of $u(r_q)$ at 
bipartition $\ms$, 
\begin{equation}\label{eq:methods_SVD1}
    \Uf(r_q) = \sum_{\alpha=1}^{d(\ms)} \lambda_{\alpha} 
    \Rs_{\alpha}(X_k) \Fs_{\alpha}(x_l)\,.
\end{equation}
This is the 1-D result corresponding to Eq.~(\ref{mpsScale}). 
For a full SVD the Schmidt number 
takes its maximal value $d(\ms)=\smaxO(\ms)$, 
where $\smaxO(\ms)=\min(2^\ms,2^{\Ns-\ms})$. 
If instead a  truncated SVD is performed by keeping only 
the $\sm(\ms) = \chi$ largest singular values, 
the error in the L$2$ norm  due to 
this approximation is $\sqrt{\sum_{\alpha=\chi+1}
^{\smaxO(\ms)} \lambda_{\alpha}^2}$.

This procedure can be straightforwardly generalised by 
replacing bits with quarternaries (2-D) or octals (3-D), 
i.e., by replacing 1-D line segments with squares (2-D) or 
cubes (3-D). The maximal Schmidt numbers are then given 
by Eq.~(\ref{gamma2d}) in 2-D and Eq.~(\ref{gamma3d}) in 3-D.

\textbf{Matrix product state algorithm.} 
The INSE are a coupled set of partial differential equations 
for the velocity field $\mf{V}$ and the pressure $p$, 
\begin{equation} \label{eq:INSE0}
\begin{split}
&\nabla \cdot \mf{V} = 0\\
&\frac{\partial\mf{V}}{\partial t} + 
\big(\mf{V} \cdot \nabla \big) 
\mf{V} = -\nabla p +  \nu \nabla^2 
\mf{V} \,,
\end{split}
\end{equation}
where $\nu$ is the kinematic viscosity and $\nabla$ is 
the nabla operator. 
After discretising the computational domain as described in the \emph{Introduction}, we solve Eq.~(\ref{eq:INSE0}) in time via a second-order Runge-Kutta method by a variational scheme. Furthermore, we use the penalty method~\cite{Fiacco1968, Fiacco1970} to satisfy the incompressibility condition $\nabla \cdot \mf{V} = 0$.  

We illustrate the principle of our method by considering 
a simple Euler time step. To advance $\mf{V}$ from 
time $t_s$ to $t_s+\Delta t$, we minimise 
the cost function 
\begin{equation}
\begin{split}\label{costfunction}
&\Theta(\mf{V}^*)=\mu\big\|
\overline{\nabla}\cdot\mf{V}^*\big\|_2^2 + \\
&\bigg\| \frac{ \mf{V}^*-\mf{V}}{\Delta t} + 
\big(\mf{V} \cdot \overline{\nabla} \big) 
\mf{V} - \nu \overline{\nabla}^2 
\mf{V} \bigg\|_2^2\,,
\end{split}
\end{equation}
where $\|\cdot\|_2$ is the L2 norm, 
$\overline{\nabla}$ is the nabla operator in finite 
difference form,   $\mf{V}^*$  is the trial 
solution at time $t_s+\Delta t$ and $\mf{V}$ denotes 
the solution at the previous time step $t_s$. 
The term  $\mu\big\|\overline{\nabla}\cdot
\mf{V}^*\big\|_2^2$ in Eq.~(\ref{costfunction}) 
enforces $\nabla \cdot \mf{V} = 0$   for sufficiently 
large values of the  penalty coefficient $\mu$. 
Note that the penalty method for enforcing the 
incompressibility condition removes the pressure $p$ 
from Eq.~(\ref{costfunction}). 
It can be calculated from the velocity fields 
via its Poisson equation~\cite{Chorin1968}. 

We represent the flow field $\mf{V}$ in terms of the 
MPS ansatz at all time steps $t_s$, and all operations on 
$\mf{V}$ like differentiation are realised via 
standard matrix product 
operators acting on the MPS~\cite[p.~591]{Lubasch2018} and~\cite[p.~22]{Ripoll2021}. 
In this way, the entire 
computation is carried out in the MPS 
manifold $\mathcal{M}\in \mathcal{D}$. A derivation of our minimisation scheme is provided in Supplementary Sec.~4.

\textbf{Direct numerical simulation algorithm}
Our DNS scheme is based upon a second-order Runge Kutta temporal discretisation combined with an eighth-order central finite difference discretisation of the spatial derivatives~\cite{Fornberg1988} on a Cartesian grid. The incompressibility condition is enforced through the projection method of Chorin \cite{Chorin1967} at every substep of each full Runge-Kutta time-step. 

The computational complexity of the DNS scheme is $\sim \Ng \log \Ng$. This is because it is dominated by the projection step, which is performed through repeated fast Fourier transform and inverse fast Fourier transforms that scale as $\sim \Ng \log \Ng$. 

If there are just enough gridpoints to resolve all scales from $l$ to $\eta$, the scheme is DNS and solves the Navier-Stokes equations \emph{exactly} within the sufficiently large $\mathcal{D}$ of Figs.~1b and 1c. If the finest scales are removed however (without invoking any subgrid scale model) such that the smallest remaining resolved scale is significantly larger than $\eta$, then the scheme becomes an URDNS operating within the scale-restricted $\mathcal{W} \subset \mathcal{D}$. The Navier-Stokes equations cannot be solved exactly within $\mathcal{W}$ due to the finest scales being subject to unphysical numerical dissipation. In comparison, the MPS algorithm operates within the MPS manifold of $\mathcal{M}$ where the interscale correlations are limited while all the scales between $l$ and $\eta$ are still present. 

\textbf{Set-up of numerical experiments.} For the  
TDJ simulations, we consider a square with edge length 
$L_{\text{box}}$ with periodic boundary conditions 
and the initial conditions 
\begin{equation}\label{eq:TDJ0}
\mf{V}(x,y,t=0) = \mf{J}(y) + \mf{D}(x,y)\,,
\end{equation} 
where $\mf{J}(y)$ is the initial jet profile
\begin{equation}\label{eq:TDJ1}
\mf{J}(y)=\hat{\mf{e}}_1 \frac{u_0}{2} 
\left[ \tanh\left(\frac{y - y_{\text{min}}}{h}\right) - \tanh\left(\frac{y 
- y_{\text{max}}}{h}\right) -1 \right]
\end{equation} 
with the streamwise direction along $\hat{\mf{e}}_1$. 
$u_0$ is the magnitude of the velocity differential between the 
jet and its surroundings, $y_{\text{min}}$ and 
$y_{\text{max}}$ describe the extent of the jet and 
$h$ is the initial thickness of the vortex sheet. 
These parameters define the Reynolds number
$\text{Re} = u_0 h/\nu$ and the time scale 
$T_0=L_{\text{box}}/u_0$.
The function 
\begin{align}
\mf{D} = \delta(\hat{\mf{e}}_1 d_1 + 
\hat{\mf{e}}_2  d_2)
\end{align}
in Eq.~(\ref{eq:TDJ0}) 
is a small disturbance of miscellaneous wave modes needed to 
initiate the roll-up of the 
jet, statistically homogeneous along 
$\hat{\mf{e}}_1$ and divergence free, with 
\begin{align}\label{eq:IncompNS2_3}
d_1(x,y) =&2\frac{L_{\text{box}}}{h^2}\left[(y-y_{\text{max}})
e^{-(y-y_{\text{max}})^2/h^2} + \right. \notag\\
&\qquad 
\left.(y-y_{\text{min}})e^{-(y-y_{\text{min}})^2/h^2} \right]
\left[\sin(8\pi x/L_{\text{box}}) +\right.
\notag\\
&\left.\qquad\quad \sin(24\pi x/L_{\text{box}}) 
+ \sin(6\pi x/L_{\text{box}})\right]\,,
\\
d_2(x,y) = &\pi
\left[e^{-(y-y_{\text{max}})^2/h^2} + e^{-
(y-y_{\text{min}})^2/h^2} \right] \times\notag \\
& \qquad \left[ 8\cos(8\pi x/L_{\text{box}}) + 
24\cos(24\pi x/L_{\text{box}})  \qquad \right. \notag \\
&\qquad \left. + 6\cos(6\pi x/L_{\text{box}})\right]\,,
\end{align}
$\delta = u_0/(40A)$ and $A = \max\limits_{x, y}\sqrt{ d_1^2 +d_2^2}$.
The components of the Reynolds stress-tensor 
shown in Fig.~\ref{fig:sims2D}b are defined as 
\begin{align}
\label{stress}
\tau_{ij}(y, t) =  \overline{u_i^{\prime}  u_j^{\prime}},
\end{align}
where $u_i^{\prime}= u_i - \overline{u}_i$ 
is the fluctuating part of the velocity 
component. 
The overbar denotes the ensemble-average across the 
statistically homogeneous streamwise direction, 
\begin{align}\label{eq:ensemAvg}
 \overline{u}_i = \frac{1}{L_{\text{box}}}
 \int\limits_0^{L_{\text{box}}} u_i(x,y,t) dx\,.
\end{align}
Scaling all lengths with $L_{\text{box}}$,  velocities with $u_0$ and time with $T_0$, we set $y_{\text{min}}= 0.4$, $y_{\text{max}}= 0.6$, $h=1/200$, and the penalty coefficient is $\mu=2.5\times10^5$ in all MPS simulations.

The TGV simulations in 3-D are conducted on a cube with edge length 
$L_{\text{box}}$ with periodic boundary conditions. We 
consider the initial flow field 
\begin{equation}\label{eq:TGV0}
\begin{split}
    &u_1(\mf{r},0) = -u_0 \sin(k_0 x) \cos(k_0 y) \cos(k_0 z)
    \,,\\
    &u_2(\mf{r},0) = u_0 \cos(k_0 x) \sin(k_0 y) \cos(k_0 z)\,,\\
    &u_3(\mf{r},0) = 0\,,\\
\end{split}
\end{equation}
where $u_0$ is the velocity amplitude of the 
initial vortex and  its wavenumber is 
$k_0 = 2\pi /L_{\text{box}}$. The corresponding 
energy at $t=0$ is $E_0= u_0^2/2$, and the 
Reynolds number is defined using the integral scale as $\text{Re} = u_0/(k_0\nu)$.

In Fig.~\ref{fig:sims3D}b we show the total kinetic energy dissipation
\begin{align}\label{eq:dedt}
\varepsilon(t) = -\frac{1}{2}
\frac{d}{dt} \int\limits_{\mathcal{V}}   
\big |\mf{V}(\mf{r},t)\big |^2 \text{d}\mathbf{r} 
\end{align}
and enstrophy
\begin{align}\label{eq:enst}
\zeta(t) = \nu \int\limits_{\mathcal{V}}   \big | \nabla \times \mf{V}(\mf{r},t) \big |^2 \text{d}\mathbf{r} ,
\end{align}
where we integrate over the whole space $\mathcal{V}$. 
$\zeta$ is related to the viscous dissipation of kinetic energy~\cite{Foias2001}. 
For incompressible flows  with periodic boundary conditions, the INSE imply 
$\varepsilon(t) = \zeta(t)$. However, restricting the 
NVPS results in numerical diffusion 
violating this equality.
In all TGV simulations, we set $T_0=L_{\text{box}}/u_0$ 
and scale lengths and velocities with $L_{\text{box}}$ 
and $u_0$, respectively. In these units, 
the dimensionless penalty coefficient 
for MPS simulations is $\mu = 6.25\times10^4$. 

\textbf{Quantitative comparison between simulations} The accuracy of the MPS and URDNS simulations are gauged by comparing the \emph{ensemble-aggregated} quantities of Figs~\ref{fig:sims2D}b and~\ref{fig:sims3D}b against DNS. Statistical quantities such as those must always be used when comparing different simulations due to the chaotic nature of turbulence. To aid the reader, we here provide a quantitative measure of the accuracies of MPS and URDNS by integrating the discrepancy DNS has to MPS and URDNS in Figs~\ref{fig:sims2D}b and~\ref{fig:sims3D}b.

We numerically calculate the \emph{visual} difference between the Fig.~\ref{fig:sims2D}b Reynolds stress of DNS (row 1) and that of MPS (rows 2-4) and URDNS (rows 5-7) as
\begin{equation}\label{eq:rms0}
 \sigma(s,c) = \sqrt{\frac{\int_{L_{\textnormal{box}}/5}^{4 L_{\textnormal{box}}/5} \int_{0}^{2T_0} \left[ \tau^{\textnormal{DNS}}_{12}(y,t) - \tau^{s,c}_{12}(y,t) \right]^2 \textnormal{d}y \textnormal{d}t}{ \frac{6}{5}L_{\textnormal{box}} T_0 \Big[\max\limits_{y, t}\big(\tau^{\textnormal{DNS}}_{12}(y,t)\big) - \min\limits_{y, t}\big(\tau^{\textnormal{DNS}}_{12}(y,t)\big) \Big]^2 }},
\end{equation}
with $s$ being the scheme in question and $c$ the compression ratio of said scheme compared to DNS (as defined in \emph{Results}). $\sigma(s,c)$ quantifies the root-mean-square of the \emph{visual} difference between the subplots in Fig.~\ref{fig:sims2D}b, and the closer $\sigma(s,c)$ is to $0$, the nearer the Reynolds stress in question is to that of DNS. $\sigma(s,c)$ is tabulated in Table~\ref{tab}.

We also quantify the accuracy of URDNS and MPS against DNS for the TGV flow by integrating the numerical diffusion $|\zeta(t)-\epsilon(t)|$ of Fig.~\ref{fig:sims3D}b. This is equivalent to
\begin{equation}\label{eq:tgvErr}
    e(s,c) = \frac{1}{E_0}\int_0^{2T_0} |\zeta_{s,c}(t)-\epsilon_{s,c}(t)| \textnormal{d} t,
\end{equation}
when normalised by the initial total kinetic energy $E_0$. The lower $e(s,c)$ is, the better is the accuracy of the relevant simulation. $e(s,c)$ is tabulated in Table~\ref{tab}. For comparison, we note that the corresponding (miniscule) error of DNS is $e(\textnormal{DNS}) = 0.002$.

\begin{table}[!h]
\centering
\begin{tabular}{c| c |c | c}
Case & Compression & Scheme & Inaccuracy \\
TDJ	& $1{:}1$ & DNS & 0 \\
\hline
TDJ	& $1{:}8$ & MPS & 0.0119 \\
\hline
TDJ & $1{:}8$ & URDNS & 0.2612 \\
\hline
TDJ	& $1{:}16$ &MPS & 0.0485 \\
\hline
TDJ & $1{:}16$ & URDNS & 0.3333 \\
\hline
TDJ & $1{:}64$ & MPS & 0.2404 \\
\hline
TDJ & $1{:}64$ & URDNS & 0.3201 \\
\hline
TGV & $1{:}1$ & DNS & 0.002 \\
\hline
TGV & $1{:}25$ & MPS & 0.0385 \\
\hline
TGV	& $1{:}25$ & URDNS & 0.1599 \\
\hline
TGV	& $1{:}49$ & MPS & 0.0844 \\
\hline
TGV & $1{:}49$ & URDNS & 0.2133 \\
\hline
TGV & $1{:}78$ & MPS & 0.2618 \\
\hline
TGV & $1{:}78$ & URDNS & 0.4563 \\
\end{tabular}
\caption{\textbf{Quantitative comparison between MPS and URDNS simulations for the TDJ and TGV flow cases}. The rows corresponding to the TDJ flow tabulate $\sigma(s,c)$, as defined in Eq.~\eqref{eq:rms0}. $\sigma(s,c)$ measures the discrepancy of the URDNS and MPS Reynolds stresses to that of DNS in Fig.~\ref{fig:sims2D}b. The TGV flow case rows tabulate $e(s,c)$, which is defined in Eq.~\eqref{eq:tgvErr} and represents the total numerical diffusion in Fig.~\ref{fig:sims3D}b. The nearer $\sigma(s,c)$ or $e(s,c)$ is to $0$, the more accurate the simulation in question is.}\label{tab}
\end{table}

\section{Data availability}
Our Code Ocean capsule~\cite{Gourianov2021} contains the raw output data from our MPS simulations. This data was generated using the \texttt{C}-functions \texttt{tntMpsBoxTurbulence2DTimeEvolutionRK2(...)} and \texttt{tntMpsBoxTurbulence3DTimeEvolutionRK2(...)}, using the initial conditions and parameters defined in the \emph{Set-up of numerical experiments} section in \emph{Methods}. Source Data for Figures~\ref{fig:CompSpaces},~\ref{fig:sims2D} and~\ref{fig:sims3D} is available with this manuscript.

\section{Code availability}
The MATLAB code required to reproduce Figs.~\ref{fig:CompSpaces},~\ref{fig:sims2D} and~\ref{fig:sims3D} is available via Code Ocean~\cite{Gourianov2021}. The MPS simulations were done using the Tensor Network Theory Library~\cite{AlAssam2017}. 

\section{Acknowledgements}
The work at the University of Oxford was supported by the EPSRC Programme Grant DesOEQ (EP/P009565/1), and MK and DJ acknowledge financial support from the National Research Foundation, Prime Ministers Office, Singapore, and the Ministry of Education, Singapore, under the Research Centres of Excellence program, and furthermore, DJ acknowledges support by the excellence cluster 'The Hamburg Centre for Ultrafast Imaging - Structure, Dynamics and Control of Matter at the Atomic Scale' of the Deutsche Forschungsgemeinschaft. Current work at the University of Pittsburgh is supported by the National Science Foundation under Grant CBET-2042918. SD is thankful for the support from EPSRC New Investigator Award (EP/T031255/1) and New Horizons grant (EP/V04771X/1). The funders had no role in study design, data collection and analysis, decision to publish or preparation of the manuscript. The authors would like to acknowledge the use of the University of Oxford Advanced Research Computing (ARC) facility in carrying out this work (\url{http://dx.doi.org/10.5281/zenodo.22558}). Finally, we thank the scientists and engineers at BAE Systems, Bristol for fruitful discussions and advice.

\section{Author contributions}
DJ conceived the research project and NG, ML, PG and DJ jointly planned it. NG, ML, SD, MK and DJ developed the quantum-inspired measure for interscale correlations based on Schmidt decompositions and hierarchical lattices. NG, ML and SD formulated the matrix product state algorithm and did the analytical calculations. NG, QYB and MK wrote the software. NG, HB and PG designed the numerical experiments for comparing MPS, URDNS and DNS. NG performed the numerical experiments. NG, ML, SD, HB, PG, MK and DJ analysed and interpreted the numerical results. NG, MK and DJ wrote the manuscript with contributions from ML, SD, HB, PG, and QYB helped revise the manuscript. The supplemental information was written by NG, ML, SD and MK. DJ supervised the project. 

\section{Competing interests}
The authors declare no competing interests.


%

\end{document}


\title{Supplementary \DIFdelbegin \DIFdel{Material}\DIFdelend \DIFaddbegin \DIFadd{Information}\DIFaddend :
A Quantum Inspired Approach to Exploit Turbulence Structures
} 

\author{Nikita Gourianov$^{1}$}
\DIFdelbegin 
\DIFdelend \DIFaddbegin \email{nikitn@hotmail.com}
\DIFaddend 

\author{Michael Lubasch$^{2}$}
\author{Sergey Dolgov$^{3}$}
\author{Quincy Y.  van den Berg$^{1}$}
\author{Hessam Babaee$^{4}$}
\author{Peyman Givi$^{4}$}
\author{Martin Kiffner$^{5,1}$}
\author{Dieter Jaksch$^{1,5,6}$}

\DIFdelbegin 
\DIFdelend \DIFaddbegin \affiliation{$^{1}$Clarendon Laboratory, University of Oxford, Oxford, UK}
\affiliation{$^{2}$Cambridge Quantum Computing Limited, London, UK}
\affiliation{$^{3}$Department of Mathematical Sciences, University of Bath, Bath, UK}
\affiliation{$^{4}$Department of Mechanical Engineering and Materials Science, University of Pittsburgh, Pittsburgh, PA, USA}
\affiliation{$^{5}$Centre for Quantum Technologies, National University of 
Singapore, Singapore}
\affiliation{$^{6}$Insitut für Laserphysik, Universität Hamburg, Hamburg, Germany}
\DIFaddend 

\DIFdelbegin 





\DIFdelend \maketitle
\DIFdelbegin 

\DIFdelend \tableofcontents            

\clearpage
\newpage
\section{Additional results}
This section expands upon the results of the main text. In particular, we provide further details on interscale correlations obtained from Schmidt decompositions of the DNS results of the TDJ (Re=1000) and TGV (Re=800) flows studied in the main text. In Sec.~\ref{subsec:SchmidtSpectra} we examine the Schmidt spectra $\lambda_\alpha$ and in Sec.~\ref{subsec:EntanEntr} we calculate the von Neumann entanglement entropy $H(n,t)$ following from these spectra.
\DIFdelbegin \DIFdel{Furthermore, to aid the reader in discriminating between the results of DNS, URDNS and MPS in the main text for the TDJ and TGV flows, we calculate the differences between the seven Reynolds stresses of Fig.~2b and integrate the numerical dissipation of Fig.~3b. The resulting scalar values are provided in the tables of Sec.~\ref{subsec:quantified}.
}\DIFdelend %
\begin{figure}[b!]
\includegraphics[width=1\linewidth]{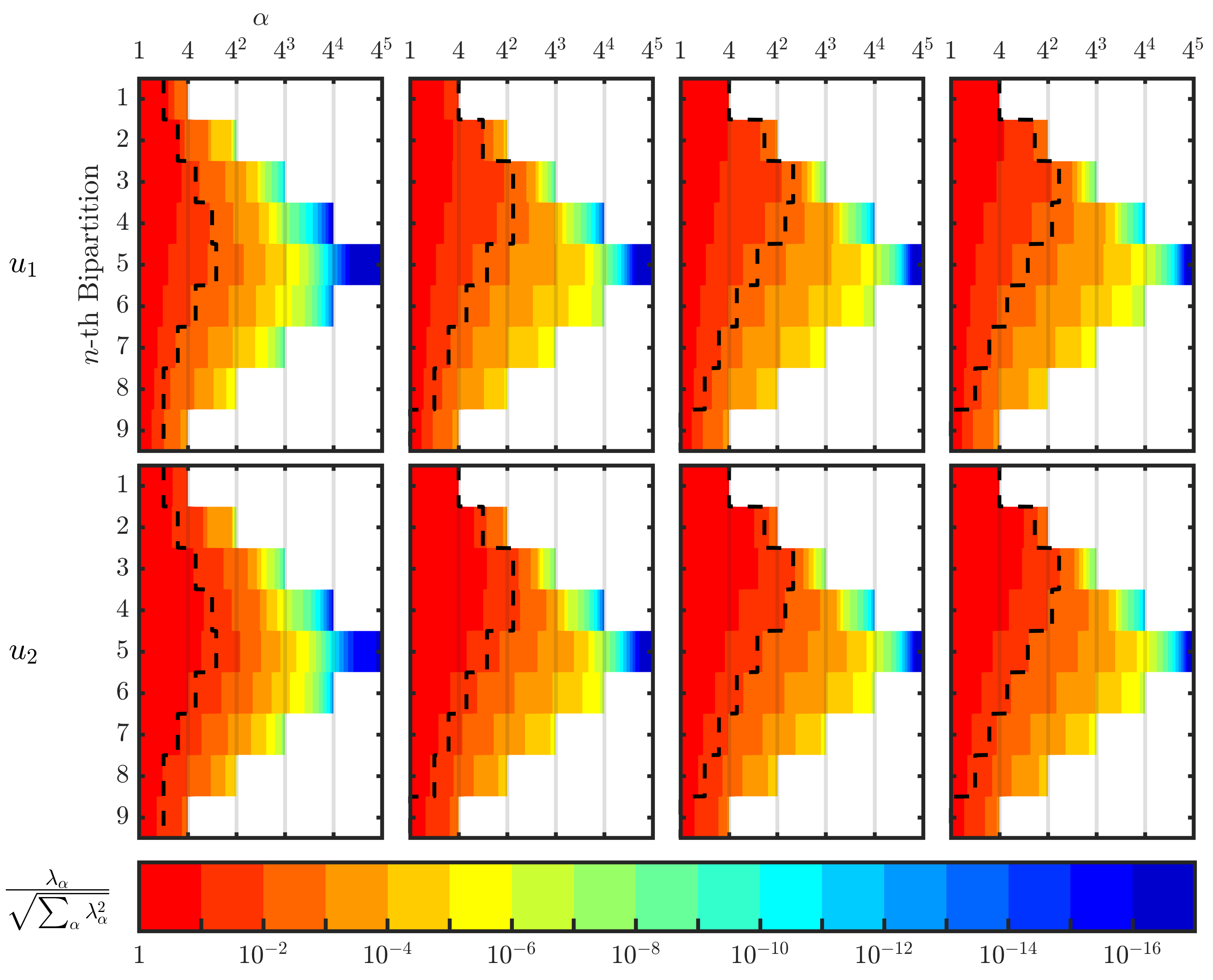}
\caption{\label{fig:S1} \textbf{Schmidt spectrum of the TDJ}. The normalised Schmidt coefficients obtained from the DNS of each velocity component of the TDJ flow at Re=1000 are shown at times $t/T_0=0.25,0.75,1.25,1.75$ (left to right), for each of the $9$ bipartitions available on the $1024\times 1024$ DNS grid. The Schmidt coefficients are sorted in descending order and are normalised such that the sum of their squares equals $1$. The black, dashed lines denote the $\SchmidtNumber_{99}(n,t)$ as used in the main text.}
\end{figure}
\subsection{Schmidt spectra}
\label{subsec:SchmidtSpectra}

\begin{figure}[t!]
\includegraphics[width=1\linewidth]{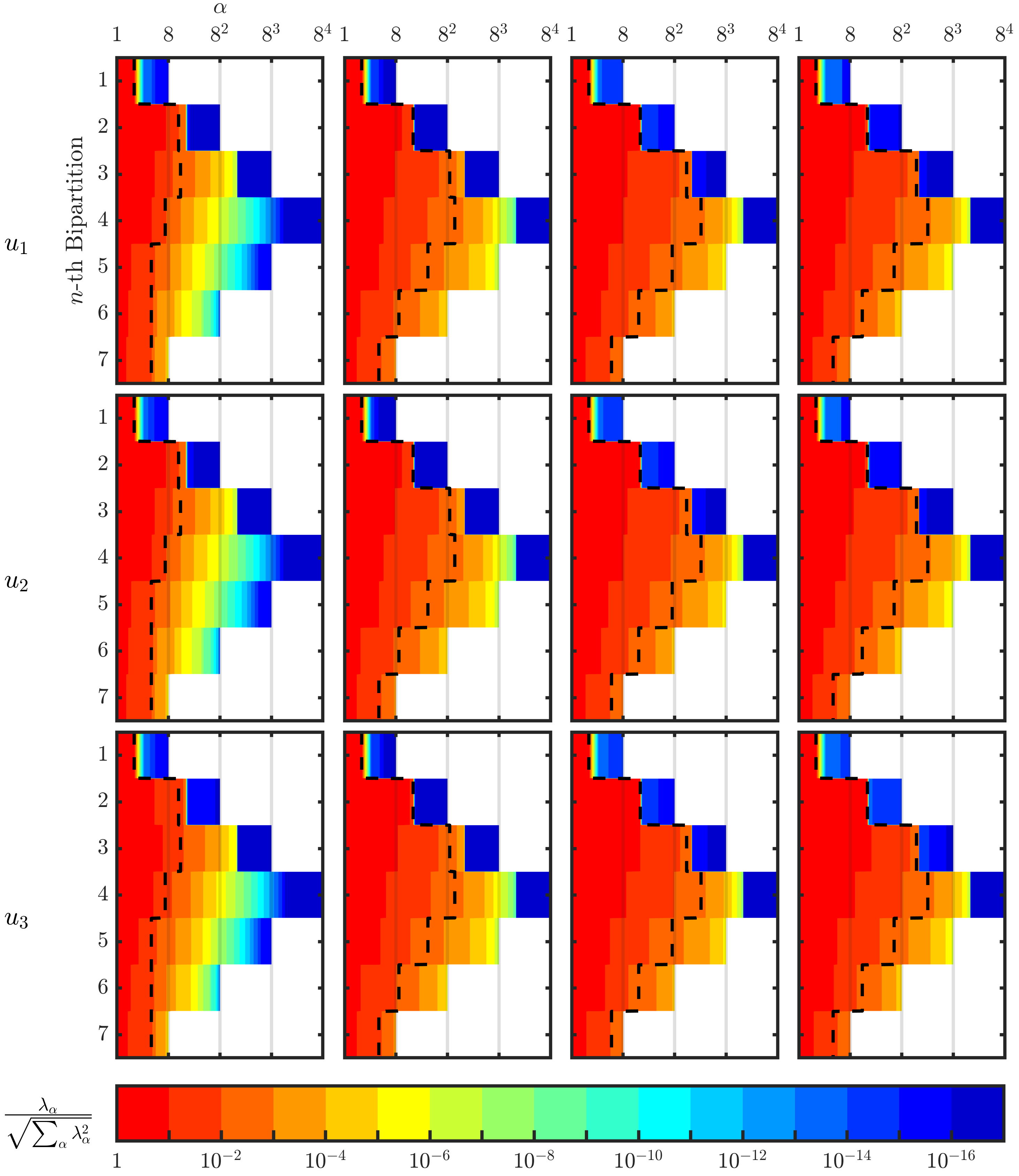}
\caption{\label{fig:S2} \textbf{Schmidt spectrum of the TGV}. The normalised Schmidt coefficients obtained from the DNS of each velocity component of the TGV flow for Re=800 are shown at times $t/T_0=0.2,0.8,1.4,2$ (left to right) for each of the $7$ bipartitions available on the $256\times256\times256$ DNS grid. The Schmidt coefficients are sorted in descending order and are normalised such that the sum of their squares equals $1$. The black, dashed lines denote the $\SchmidtNumber_{99}(n,t)$ as used in the main text.}
\end{figure}

Supp. Figs.~\ref{fig:S1} and \ref{fig:S2} show the Schmidt spectra of the 2-D TDJ and 3-D TGV flows for different times and bipartitions $n$. We have also included in both figures contours corresponding to $d_{99}(n,t)$ used as a single figure of merit for the overall accuracy of the velocity field in the main text. This shows that truncating the Schmidt spectra at these values does not discard any relevant interscale correlations of the flow.  
%
%
\subsection{Entanglement entropy}
\label{subsec:EntanEntr}
We adopt the standard definition of the von Neumann entanglement entropy $H(n,t)$ from quantum information theory. For a given spectrum $\lambda_\alpha = \lambda_\alpha(n,t)$ this is defined as
\begin{Sequation}\label{eq:EntanEntr}
H(n,t) = \frac{-1}{E(t)} \sum_{\alpha=1}^{d(n)} \lambda_{\alpha}(n,t)^2 \log \big[\lambda_{\alpha}(n,t)^2/E(t) \big],
\end{Sequation}
with the normalisation factor $E(t) = \sum_{\alpha=1}^{d(n)} \lambda_{\alpha}(n,t)^2$. 
\begin{figure}[b!]
\includegraphics[width=1\linewidth]{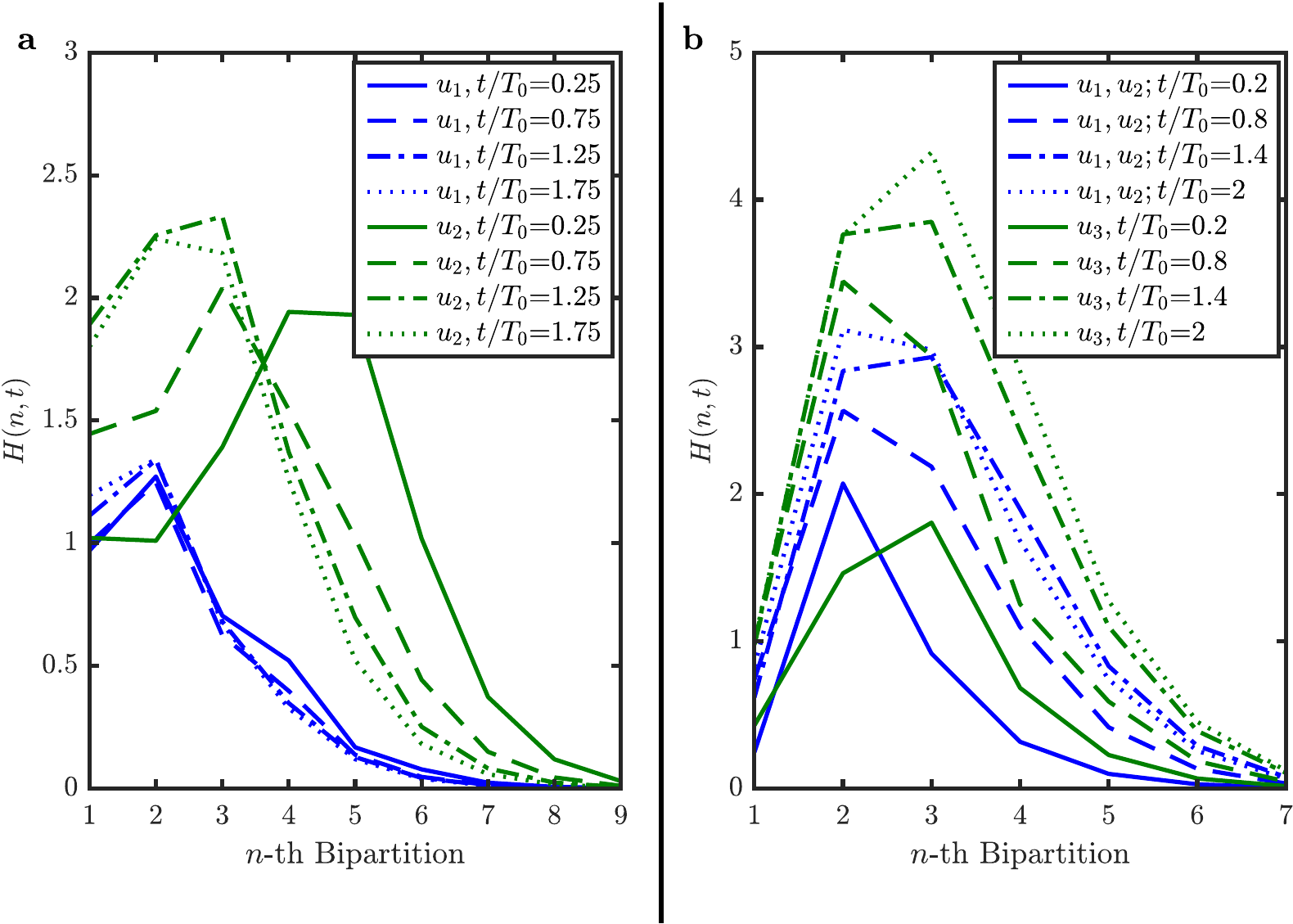}
\caption{\label{fig:S3} \textbf{Von Neumann entanglement entropy between length scales in the TDJ and TGV flows}. The entanglement entropy at each of the bipartitions of TDJ (\textbf{a}) and TGV \textbf{(b)} calculated from the DNS solutions (i.e. $d(n)$ is maximal for all $n$) are illustrated at various times. The $1024 \times 1024$ grid of the DNS of the TDJ are bipartitioned along $9$ length scales, while $7$ bipartitions are used for the $256 \times 256 \times 256$ grid DNS of the TGV. A symmetry exists between the $u_1$ and $u_2$ components of the TGV velocity field, which makes them overlap within (\textbf{b}).}
\end{figure}
The entanglement entropy for the 2-D TDJ flow shown in Supp. Fig.~\ref{fig:S3}a shifts towards bipartitions between coarser length scales (i.e. lower $n$) with increasing time. This behaviour is consistent with a 2-D inverse energy cascade~\cite{ChenShiyiETAL} where energy is carried from fine to coarse length scales as time progresses, e.g. through vortex merging. These dynamics are particularly pronounced for the cross-stream $u_2$ velocity component, for which a large number of fine scale disturbances become energised by the shear and grow in size ($t/T_0\approx 0.25,0.75$) until the eventual collapse of the jet and the saturation of the shear layer ($t/T_0\approx 1.25,1.75$). At later times no further growth of the disturbances occur. Remarkably, these physics are visible in the dynamics of the entanglement entropy. At $t/T_0=0.25$ the entanglement entropy is large for all bipartitions between $n=1$ and $n=7$, indicating significant correlations between all length scales. At later times ($t/T_0=1.25, 1.75$), when the energy increasingly flows towards coarser length scales, also the entanglement entropy shifts towards lower $n$ bipartitions as shown in Supp. Fig.~\ref{fig:S3}a.

For the 3-D TGV flow shown in Supp. Fig.~\ref{fig:S3}b the opposite happens. There, fine length scales become energised with increasing time. Correspondingly, the entanglement entropy increases at larger values of $n$ with increasing time. This increase is consistent with the hypothesis of a direct energy cascade in 3-D turbulent flows~\cite{Kolmogorov1941}, where energy is transported to progressively finer and finer length scales until the Kolmogorov microscale is reached and the energy starts being dissipated by viscosity. However, unlike in the 2-D TDJ case, the outflow of energy is not accompanied by a corresponding reduction of interscale correlations. Instead, the entanglement entropy increases with time for all bipartitions. This is a result of the disorder due to the collapse of the TGV into the various worm-like vortical structures discussed in the main text. 
We finally note that both of these behaviours for the 2-D TDJ and the 3-D TGV are consistent with the dynamics of $d_{99}$ studied in the main text.

\clearpage
\section{Matrix product state representation}\label{sec:mps}
Here we describe how the matrix product state (MPS) formalism can be used to encode scalar functions, vector fields as well as quantum wavefunctions. The encoding of scalar functions is outlined in Sec.~\ref{encoding}. A general equation is provided for the number of parameters available in MPSs in Sec.~\ref{dof}, and Sec.~\ref{schmidt} derives Eq.~(2) of the main text by performing a Schmidt decomposition in the MPS format. We describe in Sec.~\ref{EncodingVectorFields} our strategy for encoding vector fields into MPS. Finally, Secs.~\ref{subsec:encodingQuantumSystems} and~\ref{subsec:quantumSchmidtDecomp} demonstrate the MPS ansatz and Schmidt decomposition in the context of quantum mechanics for comparison.
\subsection{Encoding a flow component as a matrix product state\label{encoding}}
Consider a flow in a $K$-dimensional cube with edge length $L_{\text{box}}$ where each spatial dimension is discretised by $2^{\Ns}$ grid points. The whole $K$-dimensional grid thus comprises \DIFdelbegin \DIFdel{$2^{K\Ns}$ }\DIFdelend \DIFaddbegin \DIFadd{$\Ng=2^{K\Ns}$ }\DIFaddend equally spaced points $\mf{r}_q$. First, we introduce a one-to-one mapping between the grid point vectors $\mf{r}_q$ and a tuple of positive integers, 
\begin{Sequation}
 \mf{r}_q\leftrightarrow(q^1,q^2,\ldots,q^K)\,,
 \label{onemap}
\end{Sequation}
where $q^i\in\{0,\ldots,2^{\Ns}-1\}$ is the index of the grid point in the direction $\hat{\mf{e}}_i$. The binary representation 
$(\ldots)_2$ 
of these indices $q^i$ requires $\Ns$ bits,
\begin{Sequation}\label{eq:multi0}
 q^i=\left(\sigma^i_1,\sigma^{i}_{2},\ldots,\sigma^i_{\Ns}\right)_2\,,
\end{Sequation}
where $\sigma^i_n\in\{0,1\}$, $n=1,\ldots,\Ns,$ and  $\sigma^i_1$ and 
$\sigma^i_{\Ns}$ are the most and least significant bits, respectively.

Now consider a single Cartesian velocity 
component $u$ and ignore its time dependence.
The discretisation renders 
the velocity components into functions of the grid points $\mf{r}_q$. Due 
to the one-to-one mapping in Eq.~(\ref{onemap}), they can 
also be regarded as functions of the indices $q^i$,
$
u(\mf{r}_q) \equiv u(q^1,\ldots,q^K),
$
and hence ultimately of $\sigma^i_n$.
However, there is a freedom in the order of mapping $\sigma^i_n$ into $\mf{r}_q$.
We group all indices associated with the same bit, i.e. length scale:
\begin{Sequation}
\omega_n=\left(\sigma^1_n,\sigma^2_n,\ldots,\sigma^K_n\right)_2,
 \label{scaleOm}
\end{Sequation}
such that $\omega_n \in \{0,\ldots,2^K-1\}.$
We approximate the discrete function values $u(\mf{r}_q)$ by
an MPS $v(\mf{r}_q,\chi)$ defined as follows:
\begin{Sequation}\label{mpsF}
u(\mf{r}_q) \approx v(\mf{r}_q,  \chi) = A^{\omega_1}A^{\omega_2}\cdots 
A^{\omega_\Ns}\,,
\end{Sequation}
where the matrices $A^{\omega_n}$ have dimensions 
$\sm(n-1)\times\sm(n)$ with 
$n=1,\ldots,\Ns$, 
\begin{Sequation}
 \sm(n)=\min\left(\smaxK{K}(n),\chi\right)
\end{Sequation}
and $\smaxK{K}(n) = \min\left(2^{Kn}, 2^{K(\Ns-n)}\right)$, cf. $\smaxD(n)$ in the main text. 
Note that $\sm(0)=\sm(\Ns)=1$, and thus the MPS evaluates to 
a number. 
Each matrix $A^{\omega_n}$ in this MPS represents a particular length scale, and the maximum amount of interscale correlations that can be captured by the MPS is controlled via the bond dimension $\chi$. If $\chi$ is set to $\chi=2^{K\floor{\Ns/2}}$, the approximation in Eq.~\eqref{mpsF} becomes exact.

\subsection{Number of physical variables \label{dof}}
The number of physical parameters of $v(\mf{r}_q,\chi)$ is given by
\begin{Sequation}\label{eq:MPSdofs}
Q = 2^K \sum_{n=1} ^\Ns \SchmidtNumber(n{-}1) \SchmidtNumber(n) - \sum_{n=1}^{\Ns-1} \SchmidtNumber(n)^2.
\end{Sequation}
The first sum is the total number of parameters in Eq.~(\ref{mpsF}) while the second sum represents the intrinsic gauge degrees of freedom of the MPS format~\cite{Holtz2012}.
When $\chi$ is maximal, i.e. $\chi=2^{K\floor{\Ns/2}}$, we get \DIFdelbegin \DIFdel{$Q=2^{K\Ns}$ }\DIFdelend \DIFaddbegin \DIFadd{$Q=2^{K\Ns}=\Ng$ }\DIFaddend and that any function on the grid can be perfectly captured in MPS form.
\DIFdelbegin 

\DIFdelend %
\subsection{Schmidt decomposition \label{schmidt}}
Next we show that the MPS representation of $u$ in Eq.~(\ref{mpsF}) 
is consistent with the representation in Eq. (2) of the main text. 
To this end, we note that the gauge degrees of freedom 
allow one to bring $v$ of Eq.~(\ref{mpsF}) to 
mixed canonical form~\cite{Schollwock2011},
\begin{Sequation} 
 v(\mf{r}_q,\chi) = \sum\limits_{\alpha=1}^{\sm(\ms)}\lambda_{\alpha}
 \left[\hat{A}^{\omega_1}\cdots
 \hat{A}^{\omega_n}\right]_{\alpha}
 \left[\hat{B}^{\omega_{n+1}}\cdots
 \hat{B}^{\omega_{\Ns}}\right]_{\alpha}\,,
 \label{decompose}
\end{Sequation}
where 
$\lambda_1\geq \lambda_2... \geq \lambda_{\sm(n)}$ 
are the  Schmidt coefficients.
The matrices $\hat{A}^{\omega_n}$ and 
$\hat{B}^{\omega_n}$ have the same dimensions as 
$A^{\omega_n}$, and satisfy the relations 
\begin{Sequation}\begin{split}
 \sum\limits_{\omega_n=0}^{2^K-1} 
 \left(\hat{A}^{\omega_n}\right)^{t}\hat{A}^{\omega_n}
 =\mathds{1}\,, \\
 \sum\limits_{\omega_n=0}^{2^K-1} 
 \hat{B}^{\omega_n}\left(\hat{B}^{\omega_n}\right)^{t}
 =\mathds{1},
 \label{canonical}
\end{split}\end{Sequation}
where $(\cdot)^{t}$ denotes the matrix transpose and $\mathds{1}$ is the identity matrix.
Defining 
\begin{Sequation}\begin{split}
 \Rs_{\alpha}(\mf{X}_k) & =
 \left[\hat{A}^{\omega_1}\cdot\ldots\cdot
 \hat{A}^{\omega_n}\right]_{\alpha}\,,  \\
  \Fs_{\alpha}(\mf{x}_l) & =
 \left[\hat{B}^{\omega_{n+1}}\cdot\ldots\cdot
 \hat{B}^{\omega_\Ns}\right]_{\alpha} \,,
\end{split}\end{Sequation}
allows us to cast Eq.~(\ref{decompose}) into the form
\begin{Sequation}
  u(\mf{r}_q) = v(\mf{r}_q,\chi) = \sum\limits_{\alpha=1}^{\sm(\ms)}\lambda_{\alpha}
 \Rs_{\alpha}(\mf{X}_k)\Fs_{\alpha}(\mf{x}_l)\,,
 \label{decompose2}
\end{Sequation}
when $\chi$ is maximal, i.e. $\chi = \smaxK{K}(n)$. The grid points  $\mf{X}_k$ (coarse grid) and 
$\mf{x}_l$ (fine grid)  are defined through 
the mapping in Eq.~(\ref{onemap}) and 
the integer values 
\begin{Sequation}
\mf{X}_k: \qquad q^i = \big(\sigma^i_1,\ldots
\sigma^i_n,0,\ldots,0\big)_2, 
\end{Sequation}
and 
\begin{Sequation}
\mf{x}_l: \qquad q^i = \big(0,\ldots,0,\sigma^i_{n+1},\ldots,
\sigma^i_{\Ns}\big)_2,
\end{Sequation}
respectively. 
By definition we have $\mf{r}_q=\mf{X}_k+\mf{x}_l$,  
and Eq.~(\ref{canonical}) implies the 
orthonormality conditions 
\begin{Sequation}
\sum_k \Rs_{\alpha} (\mf{X}_k) \Rs_{\beta} (\mf{X}_k) = \sum_l 
\Fs_{\alpha}(\mf{x}_l) \Fs_{\beta}(\mf{x}_l) = \delta_{\alpha \beta}\,,
\end{Sequation}
where $\delta_{\alpha \beta}$ is the Kronecker delta. 
Writing Eq.~(\ref{decompose2}) for all three Cartesian 
velocity components and explicitly re-introducing the time 
dependence of $u_i$, $\lambda_{\alpha}$, $\Rs(\alpha)$ and $
\Fs(\alpha)$, we obtain Eq.~(2) of the main text.

\subsection{Encoding vector fields}\label{EncodingVectorFields}
Let $\mf{V}(t,\mf{r}_q)$ be the $K$-D vector field of Eq.~(1) of the main text. We now simply follow the recipe outlined in the previous Sec.~\ref{encoding} for each velocity component $u_i$ of $\mf{V}$, while again ignoring $t$. This will result in each component individually being represented as a MPS of bond dimension $\chi$, giving the MPS vector field $\mf{W}(\mf{r}_q)$:
\begin{Sequation}\label{eq:MPSvecFiel}
\mf{V}(\mf{r}_q) \approx \mf{W}(\mf{r}_q,\chi) = \sum_{i=1}^K \hat{\mf{e}}_i A^{\omega_1}_iA^{\omega_2}_i\cdots 
A^{\omega_\Ns}_i, 
\end{Sequation}
with the matrices $A^{\omega_n}_i$ being of identical dimension to the $A^{\omega_n}$ matrices of Eq.~\eqref{mpsF}.

\subsection{Encoding quantum \DIFdelbegin \DIFdel{systems}\DIFdelend \DIFaddbegin \DIFadd{states}\DIFaddend }\label{subsec:encodingQuantumSystems}
In the above sections we describe how MPS can be used to encode scalar functions and vector fields. For the sake of comparison, we here outline how MPS are traditionally used to encode 1-D quantum many-body \DIFdelbegin \DIFdel{system}\DIFdelend \DIFaddbegin \DIFadd{states}\DIFaddend .

Assume we have $\Ns$ spin-$1/2$ particles organised along a spin chain with open boundary conditions. The pure-state wavefunction of this system is given by
\begin{Sequation}\label{eq:Theory_wf0}
\ket{\Psi} = \sum_{ \{i_n\}=0,1 } {{C}}_{i_1 i_2 \ldots  i_{\Ns}} \ket{i_1} \ket{i_2} \cdots  \ket{i_{\Ns}},
\end{Sequation}
and is an element of the $\Ns$ particle Hilbert space $\mathcal{H} = \mathcal{H}_0^{\otimes \Ns}$, with $\ket{i_n} \in \mathcal{H}_0$ and $\textnormal{span}\left(\mathcal{H}_0\right) = \left\{\ket{0},\ket{1} \right\}$. The amplitude of the basis states constituting $\ket{\Psi}$ is given by the order-$\Ns$ complex tensor $C$ of dimension $2^{\Ns}$, and the wavefunction is normalised such that $\braket{\Psi|\Psi} = 1$. It is here possible to represent $\ket{0}$ and $\ket{1}$ as respectively the vectors $(1,0)^{\dagger}$ and $(0,1)^{\dagger}$, with $(\cdot)^{\dagger}$ being the conjugate-transpose. 

The indices $i_1,i_2,\ldots i_{\Ns}$ are analogous to $\omega_n$ of Eq.~\eqref{scaleOm} for $K=1$. This means $C$ can be decomposed into a MPS analogously to $u(\mf{r}_q)$:
\begin{Sequation}\label{eq:MPS_quantum}
C_{i_1 i_2 \ldots i_{\Ns}} \approx \widetilde{C}_{i_1 i_2 \ldots i_{\Ns}}(\chi) = A^{i_1}A^{i_2}\cdots A^{i_\Ns}
\end{Sequation}
where the matrices $A^{i_n}$ have dimensions 
$\sm(n-1)\times\sm(n)$ with 
$n=1,\ldots,\Ns$, 
\begin{Sequation}
 \sm(n)=\min\left(\Gamma(n),\chi\right)
\end{Sequation}
and $\Gamma(n) = \min\left(2^{n}, 2^{\Ns-n}\right)$. Hence if $\chi$ is set to $\chi=2^{\floor{\Ns/2}}$, the relationship in Eq.~\eqref{eq:MPS_quantum} becomes exact. Due to the open boundary conditions, $\sm(0)=\sm(D)=1$ and thus the MPS evaluates to 
a number. For a detailed step-by-step guide on how the MPS decomposition can be performed using SVDs, see~\cite{Schollwock2011}. Each matrix $A^{i_n}$ of the above MPS is associated with a particular spin-$1/2$ particle, and the maximum amount of entanglement of the system that can be captured by the MPS is controlled by the bond dimension $\chi$. Compare these properties with those of the MPS of the scalar function in Sec.~\ref{encoding} for $K=1$.

The above MPS representation of $C$ allows us to rewrite the wavefunction as 
\begin{Sequation}\label{eq:wf1}
\ket{\Psi}\approx\ket{\widetilde{\Psi}(\chi)} = \sum_{ \{i_n\}=0,1 } A^{i_1}A^{i_2}\cdots A^{i_\Ns} \ket{i_1} \ket{i_2} \cdots  \ket{i_{\Ns}}.
\end{Sequation}
$\ket{\widetilde{\Psi}(\chi)}$ is still a vector within the exponentially large $\mathcal{H}$, but the MPS decomposition has parameterised it with only a \emph{polynomial} number of variables (when $\chi$ is limited). For many quantum systems, in particular area-law following ones, using the MPS decomposition does not cause a significant loss in accuracy~\cite{Orus2014}.

\subsection{Quantum Schmidt decomposition}\label{subsec:quantumSchmidtDecomp}
The Schmidt decomposition can be employed on $\ket{\widetilde{\Psi}(\chi)}$ just as it was employed on $u(\mf{r}_q,\chi)$ in Sec.~\ref{schmidt}. Putting the MPS into mixed canonical form with the canonical centre at the $n$-th bond yields 
\begin{Sequation} 
 \ket{\widetilde{\Psi}(\chi)} = \sum\limits_{\alpha=1}^{\sm(\ms)}\lambda_{\alpha}
 \left[\hat{A}^{i_1}\cdots\hat{A}^{i_n}\right]_{\alpha}
 \left[\hat{B}^{i_{n+1}}\cdots \hat{B}^{i_{\Ns}}\right]_{\alpha} \ket{i_1} \ket{i_2} \cdots  \ket{i_{\Ns}}.
 \label{decompose_quantum}
\end{Sequation}
This is nothing else than the Schmidt decomposition, as can straightforwardly be seen by inserting 
\begin{Sequation}\begin{split}
 \ket{\psi^{1:n}_{\alpha}} & =
  \left[\hat{A}^{i_1}\cdots\hat{A}^{i_n}\right]_{\alpha} \ket{i_1} \cdots \ket{i_{\Ns}}\,, \\
  \ket{\psi^{n+1:\Ns}_{\alpha}} & =
 \left[\hat{B}^{i_{n+1}}\cdots \hat{B}^{i_{\Ns}}\right]_{\alpha} \ket{i_{n+1}} \cdots  \ket{i_{\Ns}} \,
\end{split}\end{Sequation}
into Eq.~\eqref{decompose_quantum}, yielding the Schmidt decomposition between two quantum sub-systems in its traditional form:

\begin{Sequation} 
 \ket{\widetilde{\Psi}(\chi)} = \sum\limits_{\alpha=1}^{\sm(\ms)}\lambda_{\alpha}
 \ket{\psi^{1:n}_{\alpha}}
  \ket{\psi^{n+1:\Ns}_{\alpha}}.
 \label{decompose_quantum_2}
\end{Sequation}
The difference between this Schmidt decomposition and the one in Eq.~\eqref{decompose}, is that the Schmidt decomposition of Eq.~\eqref{decompose} exposes the interscale correlations between a set of coarse and fine length scales, whilst the quantum Schmidt decomposition reveals the entanglement between two bipartitions of the spin chain

For a detailed description of how the entanglement of a quantum system might be studied in the context of MPS, see~\cite{Schollwock2011}. The tools provided there can also be used to study the interscale correlations of turbulent flows, like how we calculated the von Neumann entanglement entropy of the TDJ and TGV in Sec.~\ref{subsec:EntanEntr}.

\newpage
\clearpage
\section{Matrix product state representation of shock waves}
In the following, we analyse how well MPS approximate hump-solutions of the one-dimensional Burgers' equation. First we introduce Burgers' equation and present its known mathematical solution in Sec.~\ref{subsec:BurgersEquation}.
Then in Sec.~\ref{subsec:InitialDeltaFunction} we derive the solution of Burgers' equation for an initial $\delta$ function.
We show that this so-called hump-solution has, in general, an exponentially accurate MPS representation in Sec.~\ref{subsec:MPSForInitialDeltaFunction}. In the special case of vanishing viscosity, this hump solution becomes a triangular wave, i.e. a prototypical shock-wave. We provide the exact MPS description for this shock-wave in
Sec.~\ref{subsec:MPSForTriangleFunction}.

\subsection{Burgers' equation and its analytical solution}
\label{subsec:BurgersEquation}

Burgers' equation~\cite{Bateman1915, Burgers1948} is given by
\begin{Sequation}\begin{split}\label{eq:BurgersEquation}
 \frac{\partial u}{\partial t} & = \nu \frac{\partial^{2} u}{\partial x^{2}} - u \frac{\partial u}{\partial x}
\end{split}\end{Sequation}
where $u = u(x, t)$.
This equation simplifies by defining a function $w = w(x, t)$ via
\begin{Sequation}\begin{split}\label{eq:Theta}
 u(x, t) & = -2 \nu \frac{1}{w} \frac{\partial w}{\partial x} \\
           & = -2 \nu \frac{\partial \log(w)}{\partial x} ,
\end{split}\end{Sequation}
also known as the Hopf-Cole transformation~\cite{Hopf1950, Cole1951}, which leads to the heat equation
\begin{Sequation}\begin{split}\label{eq:HeatEquation}
 \frac{\partial w}{\partial t} & = \nu \frac{\partial^{2} w}{\partial x^{2}} .
\end{split}\end{Sequation}
The heat equation has the general mathematical solution
\begin{Sequation}\begin{split}\label{eq:HeatEquationSolution}
 w(x, t) & = \frac{1}{2 \sqrt{\pi \nu t}} \int_{-\infty}^{\infty} w_{0}(\alpha) e^{-\frac{(x-\alpha)^{2}}{4 \nu t}} \text{d}\alpha 
\end{split}\end{Sequation}
where $w_{0}$ denotes the initial function at time $t = 0$.
We obtain $w_{0}$ from the initial function to Burgers' equation $u_{0}(x) = u(x, 0)$ by inverting Eq.~\eqref{eq:Theta} for $t = 0$:
\begin{Sequation}\begin{split}\label{eq:InitialFunction}
 w_{0}(x) & = w(x, 0) \nonumber\\
       & = e^{-\frac{1}{2 \nu} \int_{a}^{x} u_{0}(y) \text{d}y}
\end{split}\end{Sequation}
where $a$ can be chosen freely.
Using Eq.~\eqref{eq:Theta} we obtain the following solution to Burgers' equation
\begin{Sequation}\begin{split}\label{eq:BurgersEquationSolution}
 u(x, t) & = \frac{1}{t} \frac{\int_{-\infty}^{\infty} w_{0}(\alpha) (x-\alpha) e^{-\frac{(x-\alpha)^{2}}{4 \nu t}} \text{d}\alpha}{\int_{-\infty}^{\infty} w_{0}(\alpha) e^{-\frac{(x-\alpha)^{2}}{4 \nu t}} \text{d}\alpha}
\end{split}\end{Sequation}
where $w_{0}(\alpha)$ represents the initial function defined in Eq.~\eqref{eq:InitialFunction}.

\subsection{Mathematical solution for initial $\delta$ function}
\label{subsec:InitialDeltaFunction}

Following~\cite{Whitham1974}, we investigate the so-called hump solution to the Burgers' equation. It is produced when using the $\delta$ function initial condition of
\begin{Sequation}\begin{split}\label{eq:InitialDeltaForBurgers}
 u_{0}(x) & = Z \delta(x - x_{0}),
\end{split}\end{Sequation}
where $Z$ is a normalising constant.
Plugging this into Eq.~\eqref{eq:InitialFunction} gives
\begin{Sequation}\begin{split}
 w_{0}(x) & = e^{-\frac{Z}{2 \nu} \int_{a}^{x} \delta(y - x_{0}) \text{d}y} .
\end{split}\end{Sequation}
We choose $a = x_{0} + \epsilon$, where $\epsilon \to 0$ denotes an infinitesimally small number, so that
\begin{Sequation}\begin{split}\label{eq:InitialFunctionForHeat}
 w_{0}(x) & = e^{\frac{Z}{2 \nu}} \quad \, x \leq x_{0},\\
 w_{0}(x) & = 1 \quad \, x > x_{0} .
\end{split}\end{Sequation}
We split the integration into two parts
\begin{Sequation}\begin{split}\label{eq:SplitIntegral}
 \int_{-\infty}^{\infty} w_{0}(\alpha) \ldots \text{d}\alpha & = \int_{-\infty}^{x_{0}} e^{\frac{Z}{2 \nu}} \ldots \text{d}\alpha + \int_{x_{0}+\epsilon}^{\infty} \ldots \text{d}\alpha.
\end{split}\end{Sequation}
To simplify our notation, we do not explicitly include $\epsilon$ from now on.
The evaluation of the integral in the numerator of Eq.~\eqref{eq:BurgersEquationSolution} is straightforward and to evaluate the integral in the the denominator we make use of the substitution $\beta = (x - \alpha) / (2 \sqrt{\nu t})$.
Ultimately these calculations lead to the result
\begin{Sequation}\begin{split}\label{eq:BurgersSolutionDelta}
 u(x, t) & = \sqrt{\frac{\nu}{t}} \frac{( e^{\frac{Z}{2 \nu}} - 1 ) e^{-\frac{(x - x_{0})^{2}}{4 \nu t}}}{\sqrt{\pi} + \frac{\sqrt{\pi}}{2} (e^{\frac{Z}{2 \nu}} - 1) \text{erfc}(\frac{x - x_{0}}{2 \sqrt{\nu t}})},
\end{split}\end{Sequation}
where $\text{erfc}(x) = (2 / \sqrt{\pi}) \int_{x}^{\infty} \exp(-\alpha^{2}) \text{d}\alpha$ is the so-called complementary error function.

To analyse the solution~\eqref{eq:BurgersSolutionDelta} in the limit of $\nu \to 0$, we first consider $x \leq x_{0}$ for which
\begin{Sequation}\begin{split}
 \lim_{\nu \to 0} \text{erfc}\bigg(\frac{x - x_{0}}{2 \sqrt{\nu t}}\bigg) & = 2, \quad \, x < x_{0}\\
 \lim_{\nu \to 0} \text{erfc}\bigg(\frac{x - x_{0}}{2 \sqrt{\nu t}}\bigg) & = 1, \quad \text{for}\, x = x_{0}
\end{split}\end{Sequation}
and therefore Eq.~\eqref{eq:BurgersSolutionDelta} becomes
\begin{Sequation}\begin{split}
 \lim_{\nu \to 0} u(x, t) & = 0, \quad \, x \leq x_{0} .
\end{split}\end{Sequation}
For $x > x_{0}$ we use the ${x \to \infty}$ asymptotic expansion of the complementary error function
\begin{Sequation}\begin{split}
\text{erfc}(x) & = \frac{e^{-x^{2}}}{\sqrt{\pi} x} + O(x^{-3} e^{-x^2})
\end{split}\end{Sequation}
so that for $\nu \to 0$
\begin{Sequation}\begin{split}
\text{erfc}\bigg(\frac{x - x_{0}}{2 \sqrt{\nu t}}\bigg) \approx 2 \sqrt{\nu t} \frac{e^{-\frac{(x - x_{0})^{2}}{4 \nu t}}}{\sqrt{\pi}(x - x_{0})}, \quad \, x > x_{0},
\end{split}\end{Sequation}
which transforms Eq.~\eqref{eq:BurgersSolutionDelta} into
\begin{Sequation}\begin{split}\label{eq:BurgersSolutionDeltaNu0Temp}
 u(x, t) & = \sqrt{\frac{\nu}{t}} \frac{( e^{\frac{Z}{2 \nu}} - 1 ) e^{-\frac{(x - x_{0})^{2}}{4 \nu t}}}{\sqrt{\pi} + \sqrt{\nu t} (e^{\frac{Z}{2 \nu}} - 1) \frac{e^{-\frac{(x - x_{0})^{2}}{4 \nu t}}}{x - x_{0}}}.
\end{split}\end{Sequation}
We observe that
\begin{Sequation}\begin{split}\label{eq:limitSols}
 \lim_{\nu \to 0} e^{-\frac{(x - x_{0})^{2}}{4 \nu t}} & = 0, \quad \, x > x_{0},\\
 \lim_{\nu \to 0} \left( e^{\frac{Z}{2 \nu} - \frac{(x - x_{0})^{2}}{4 \nu t}} \right) & = \infty, \quad \, x_{0} < x < x_{0} + \sqrt{2 Z t},\\
 \lim_{\nu \to 0} \left( e^{\frac{Z}{2 \nu} - \frac{(x - x_{0})^{2}}{4 \nu t}} \right) & = 1, \quad \text{for}\, x = x_{0} + \sqrt{2 Z t},\\
 \lim_{\nu \to 0} \left( e^{\frac{Z}{2 \nu} - \frac{(x - x_{0})^{2}}{4 \nu t}} \right) & = 0, \quad \, x > x_{0} + \sqrt{2 Z t} .
\end{split}\end{Sequation}
Using these solutions along with Eq.~\eqref{eq:BurgersSolutionDeltaNu0Temp} gives in the limit $\nu \to 0$:
\begin{Sequation}\begin{split}\label{eq:BurgersSolutionDeltaNu0}
 u(x, t) & = 0, \quad \, x \leq x_{0} \land x > x_{0} + \sqrt{2 Z t},\\
 u(x, t) & = \frac{x - x_{0}}{t}, \quad \, x_{0} < x < x_{0} + \sqrt{2 Z t} .
\end{split}\end{Sequation}

As $\nu$ grows large, it is straightforward to show that Eq.~\eqref{eq:BurgersSolutionDelta} approaches the Gaussian
\begin{Sequation}\begin{split}\label{eq:BurgersSolutionDeltaNuInfty}
 u(x, t) & = \frac{Z}{2 \sqrt{\pi \nu t}} e^{-\frac{(x - x_{0})^{2}}{4 \nu t}} .
\end{split}\end{Sequation}
And for $\nu\rightarrow \infty$, this Gaussian will approach a uniform function with an amplitude tending towards zero.

\subsection{Matrix product state representation of general solution}
\label{subsec:MPSForInitialDeltaFunction}
Here we show that MPS accurately represent the previously derived mathematical solution for the propagation of an initial $\delta$ function with Burgers' equation.
For $K=1$ the MPS digit of Eq.~\eqref{scaleOm} collapses to the range $\omega_n\equiv \sigma^1_n \in \{0,1\}$.
In turn, every grid point can be defined as $\mf{r}_q = x_q = q/2^\Ns$, where $q=\left(\omega_1,\ldots,\omega_\Ns\right)_2$.

The discretised $\delta$-function $u_0(x_q) = Z \delta(q - j)$ with a normalising constant $Z$ and a peak position $j = (b_1,\ldots,b_\Ns)_2$, $b_n \in \{0,1\}$, can be represented as an exact MPS (see Eq.~\eqref{mpsF}) of bond dimension $\chi = 1$ and factors
\begin{Sequation}
A^{\omega_1} = \left\{\begin{array}{ll} Z, & \omega_1=b_1, \\ 0, & \mbox{otherwise}, \end{array}\right. 
\qquad \mbox{and} \qquad
A^{\omega_n} = \left\{\begin{array}{ll} 1, & \omega_n=b_n, \\ 0, & \mbox{otherwise}, \end{array}\right. 
\end{Sequation}
for $n=2,\ldots,\Ns$.

Time evolution of an initial $\delta$ function with Burgers' equation has the solution of Eq.~\eqref{eq:BurgersSolutionDelta}.
The solution for $\nu \to 0$ is a triangular wave which has an exact MPS representation of bond dimension $\chi = 3$, as shown in the following section. The solution for very large $\nu$ is a Gaussian (tending towards a uniform function for $\nu\rightarrow \infty)$, which has an exponentially convergent MPS approximation~\cite{DKhOs2012}. For $0 < \nu < \infty$ it can be shown that the function of Eq.~\eqref{eq:BurgersEquationSolution} remains holomorphic in $x$, and hence regularity arguments similar to those in \cite{Herrmann2020} can be used to prove an exponential convergence of the polynomial approximation of Eq.~\eqref{eq:BurgersEquationSolution}. In turn, polynomials of degree $p$ sampled on an equidistant grid admit an exact MPS representation~\cite[Thm.~6]{Osel2013} with $\chi \le p+1$. Therefore, the $0 < \nu < \infty$ MPS approximation of Eq.~\eqref{eq:BurgersEquationSolution} is also exponentially converging.

Therefore we conclude that the MPS description of solutions of the initial values problem considered here are exponentially convergent in the number of variables used. In other words, an MPS scheme would require exponentially fewer variables than e.g. a standard finite differences scheme. For the sake of concreteness, we illustrate this for the case of $\nu\rightarrow 0$ in the following section.

\subsection{Triangular waves as matrix product states}
\label{subsec:MPSForTriangleFunction}
Let us analytically derive the MPS representation for the $\nu\rightarrow 0$ triangular-wave solution of Eq.~\eqref{eq:BurgersSolutionDelta}. This limit solution is a prototypical shock-wave and (as all shock-waves) it is discontinuous, which slows down the convergence of both polynomial and Fourier approximations. In contrast, an MPS with a bond dimension of just 3 can represent this function exactly, as we now demonstrate in this section.

\noindent \textbf{Definition.} A Heaviside vector of length $J \in \mathbb{N}$ with step position $j \in \mathbb{Z}$ is defined element-wise as
\begin{Sequation}\label{eq:heav-d}
    \theta^j_q := \left\{\begin{array}{ll}1, & q\le j, \\ 0, & \mbox{otherwise,}\end{array}\right.
\end{Sequation}
for $q \in \{0,\ldots,J-1\}$.

\noindent\textbf{Definition.} A unit vector of length $J \in \mathbb{N}$ at position $j \in \mathbb{Z}$ is defined element-wise as 
\begin{Sequation}
e^j_q := \left\{\begin{array}{ll} 1, & q=j, \\ 0, & \mbox{otherwise,}\end{array}\right.
\end{Sequation}
for $q \in \{0,\ldots,J-1\}.$

\noindent\textbf{Lemma.} Let $j=(b_1,\ldots,b_{\Ns})_2$ and $q=(\omega_1,\ldots,\omega_{\Ns})_2$, with $b_n, \omega_n \in \{0,1\}$. Then the $j$-th Heaviside vector of length $J=2^{\Ns}$ can be written as the MPS
\begin{Sequation}
\theta^{b_1 \ldots b_{\Ns}}_{\omega_1\ldots \omega_{\Ns}} = T^{\omega_1} \cdots T^{\omega_{\Ns}}
\end{Sequation}
with bond dimensions $d(n)=2$, $n=1,\ldots,{\Ns}-1$, where
\begin{Sequation}
    T^{\omega_1} = \begin{bmatrix}e^{b_1}_{\omega_1} & \theta^{b_1-1}_{\omega_1}\end{bmatrix},
\end{Sequation}
\begin{Sequation}
    T^{\omega_n} = \begin{bmatrix}e^{b_n}_{\omega_n} & \theta^{b_n-1}_{\omega_n} \\  0 & 1\end{bmatrix},
\end{Sequation}
for  $n=2,\ldots,{\Ns}-1$, and
\begin{Sequation}
    T^{\omega_{\Ns}} = \begin{bmatrix}\theta^{b_{\Ns}}_{\omega_{\Ns}} \\ 1\end{bmatrix}.
\end{Sequation}

\noindent\textbf{Proof.} Consider two indices first, and prove that $\theta^{b_1,b_2}_{\omega_1 \omega_2} = \theta^{b_1-1}_{\omega_1}  + \theta^{b_2}_{\omega_2}  e^{b_1}_{\omega_1}$.

\begin{itemize}
\item If $\omega_1>b_1$, we obtain $0$, as expected from~\eqref{eq:heav-d}.

\item If $\omega_1=b_1$, we get $\theta^{b_2}_{\omega_2}$.
This becomes $0$ when $\omega_2>b_2$, and $1$ when $\omega_2\le b_2$, which, together with $\omega_1=b_1$, gives $q = 2\omega_1 + \omega_2 \le 2 b_1 + b_2 = j$, as expected.

\item If $\omega_1 \le b_1-1$, we are left with $\theta^{b_1-1}_{\omega_1} = 1$, 
but in this case we also have that $q=2\omega_1 + \omega_2 < 2b_1 + b_2=j$.
\end{itemize}
All cases are thus in agreement with~\eqref{eq:heav-d}. Multiplying the last two factors of the Heaviside MPS gives
\begin{Sequation}
T^{\omega_{{\Ns}-1}} T^{\omega_{\Ns}} = \begin{bmatrix}\theta^{b_{{\Ns}-1}-1}_{\omega_{{\Ns}-1}} + \theta^{b_{\Ns}}_{\omega_{\Ns}} e^{b_{{\Ns}-1}}_{\omega_{{\Ns}-1}} \\ 1\end{bmatrix} = \begin{bmatrix}\theta^{b_{{\Ns}-1},b_{\Ns}}_{\omega_{{\Ns}-1}, \omega_{\Ns}} \\ 1 \end{bmatrix}.
\end{Sequation}
Similarly, assuming that
\begin{Sequation}
T^{\omega_n} \cdots T^{\omega_{\Ns}} = \begin{bmatrix}\theta^{b_n\ldots b_{\Ns}}_{\omega_n \ldots \omega_{\Ns}} \\ 1 \end{bmatrix},
\end{Sequation}
gives the induction step for $T^{\omega_{n-1}} \cdots T^{\omega_{\Ns}}$, and eventually, since $T^{\omega_1}$ is just one row,
\begin{Sequation}
T^{\omega_1} \cdots T^{\omega_{\Ns}} = \theta^{b_1 \ldots b_{\Ns}}_{\omega_1 \ldots \omega_{\Ns}}
\end{Sequation}
as expected.

\noindent\textbf{Definition.} A vector whose elements are $x_q = q = (\omega_1,\ldots,\omega_n)_2$ can be expressed by the MPS
\begin{Sequation}
X_q \equiv X_{\omega_1 \ldots \omega_{\Ns}} = \begin{bmatrix}1 & 2^{{\Ns}-1} \omega_1\end{bmatrix}  \cdots \begin{bmatrix}1 & 2^{{\Ns}-n} \omega_n \\ 0 & 1\end{bmatrix} \cdots \begin{bmatrix}\omega_{\Ns} \\ 1\end{bmatrix}
\end{Sequation}
of bond dimension $2$, per~\cite{Osel2013}.

\noindent\textbf{Theorem.} A triangular wave vector with elements
\begin{Sequation}
w^{b_1 \ldots b_{\Ns}}_{\omega_1 \ldots \omega_{\Ns}} := X_{\omega_1 \ldots \omega_{\Ns}} \cdot  \theta^{b_1 \ldots b_{\Ns}}_{\omega_1 \ldots \omega_{\Ns}}
\end{Sequation}
can be written as an MPS of bond dimension $3$.

\noindent\textbf{Proof.}
Multiplying the MPS $X_{q_1 \ldots q_{\Ns}}$ with $\theta^{b_1 \ldots b_{\Ns}}_{q_1 \ldots q_{\Ns}}$ tensor by tensor will result in an MPS $\hat W^{\omega_1} \cdots \hat W^{\omega_{\Ns}}$ of bond dimension $4$. However, this decomposition is redundant. For example, the first factor reads
\begin{Sequation}
\hat W^{\omega_1} = \begin{bmatrix}e^{b_1}_{\omega_1} & e^{b_1}_{\omega_1} 2^{{\Ns}-1} \omega_1 & \theta^{b_1-1}_{\omega_1} & \theta^{b_1-1}_{\omega_1} 2^{{\Ns}-1} \omega_1\end{bmatrix},
\end{Sequation}
albeit $e^{b_1}_{\omega_1} \omega_1 = b_1 e^{b_1}_{\omega_1}$ and this means that
\begin{Sequation}
\hat W^{\omega_1} =  \underbrace{\begin{bmatrix}e^{b_1}_{\omega_1}  & \theta^{b_1-1}_{\omega_1} & \theta^{b_1-1}_{\omega_1} 2^{{\Ns}-1} \omega_1\end{bmatrix}}_{W^{\omega_1}}  
\underbrace{\begin{bmatrix}1 & b_1 2^{{\Ns}-1} & 0 & 0 \\ 0 & 0 & 1 & 0 \\ 0 & 0 & 0 & 1\end{bmatrix}}_{R^1}.
\end{Sequation}
$W^{\omega_1}$ (with bond dimension $d(1)=3$) can be considered the first non-redundant MPS tensor of $w^{b_1 \ldots b_{\Ns}}_{\omega_1 \ldots \omega_{\Ns}}$. Multiplying $R^1$ with $\hat W^{\omega_2}$ will continue the reduction and produce $W^{\omega_2}$. Assuming that $W^{\omega_1}, \ldots, W^{\omega_{n-1}}$ have already been obtained, the next step gives
\begin{Sequation}
R^{n-1} \hat W^{\omega_n} := 
\begin{bmatrix}
1 & c_{n-1} & 0 & 0 \\ 
0 & 0       & 1 & 0  \\ 
0 & 0       & 0 & 1\end{bmatrix}
\begin{bmatrix}
e^{b_n}_{\omega_n}  &  2^{{\Ns}-n}\omega_n e^{b_n}_{\omega_n} && \theta^{b_n-1}_{\omega_n} & 2^{{\Ns}-n} \omega_n \theta^{b_n-1}_{\omega_n} \\
0                   &  e^{b_n}_{\omega_n}                 &&      0                    &  \theta^{b_n-1}_{\omega_n} \\\\  
0                   &  0                                  &&      1                    &  2^{{\Ns}-n} \omega_n \\ 
0                   &  0                                  &&      0                    & 1 \end{bmatrix},
\end{Sequation}
where $c_{n-1}$ is a scalar, with $c_1=b_1 2^{{\Ns}-1}$. This gives
\begin{Sequation}
R^{n-1} \hat W^{\omega_n} = \begin{bmatrix}
e^{b_n}_{\omega_n}  & (2^{{\Ns}-n}b_n + c_{n-1}) e^{b_n}_{\omega_n} && \theta^{b_n-1}_{\omega_n} &  (2^{{\Ns}-n} \omega_n + c_{n-1}) \theta^{b_n-1}_{\omega_n} \\
0                   & 0                                         && 1                         &  2^{{\Ns}-n} \omega_n \\
0                   & 0                                         && 0                         &  1
\end{bmatrix}.
\end{Sequation}
Notice that the first two columns are linearly dependent, allowing us to rewrite the above expression into
\begin{Sequation}
\begin{split}
&R^{n-1} \hat W^{\omega_n} =\\
&\underbrace{\begin{bmatrix}
e^{b_n}_{\omega_n}  &  \theta^{b_n-1}_{\omega_n} &  (2^{{\Ns}-n} \omega_n + c_{n-1}) \theta^{b_n-1}_{\omega_n} \\
0                   &  1                         &  2^{{\Ns}-n} \omega_n \\
0                   &  0                         &  1
\end{bmatrix}}_{W^{\omega_n}}
\underbrace{\begin{bmatrix}
1   & 2^{{\Ns}-n}b_n + c_{n-1}  && 0  & 0 \\
0   & 0                     && 1  & 0 \\
0   & 0                     && 0  & 1 
\end{bmatrix}}_{R^n}.\\
\end{split}
\end{Sequation}
Since $R^n$ is identical to $R^{n-1}$ except for the element $c_n=2^{{\Ns}-n}b_n + c_{n-1}$ at position $(1,2)$, the recursion can continue all the way until $\hat W^{\omega_{\Ns}}$. This will result in the factors $W^{\omega_n}$ whose bond dimensions are all $d(n) = 3$, and these factors together form the non-redundant representation of $w^{b_1 \ldots b_{\Ns}}_{\omega_1 \ldots \omega_{\Ns}}$. 

\newpage
\clearpage
\section{Matrix product state algorithm}
In this section we present our MPS algorithm for minimising the cost function of Eq.~(8) in the main text in 2-D (i.e. $K=2$). Generalising this scheme to other dimensions is straightforward. Sec.~\ref{secMPS:A} explains how Eq.~(8) can be considered a variational problem to be solved within the MPS manifold. The minimisation can be done by repeatedly minimising for each of the matrices of the MPS representing the variational flow field. Our procedure for this local minimisation is outlined in Sec.~\ref{secMPS:B}. Sec.~\ref{secMPS:C} explains how repeatedly performing many such local minimisations allows us to converge to the global minimum of the cost function. The full computational complexity of our algorithm is derived in Sec.~\ref{secMPS:D}. The computational complexity is also demonstrated in practice in Sec.~\ref{secMPS:E}. Finally, in Sec.~\ref{MPS:F} we discuss the arithmetic intensity of the MPS algorithm. 

\subsection{The variational problem}\label{secMPS:A}
Let us begin by expanding upon main text Eq.~(8). In 2-D, the finite difference del operator is given as \begin{Sequation}
\overline{\nabla} = \hat{\mf{e}}_1 \frac{\Delta}{\Delta x_1} + \hat{\mf{e}}_2 \frac{\Delta}{\Delta x_2},
\end{Sequation} with $\frac{\Delta}{\Delta x_k}$ being the derivative along unit vector $\hat{\mf{e}}_k$, and the Laplace operator by 
\begin{Sequation}
\overline{\nabla}^2 = \frac{\Delta^2}{\Delta x_1^2} + \frac{\Delta^2}{\Delta x_2^2},
\end{Sequation}
with $\frac{\Delta^2}{\Delta x_k^2}$ the second derivative along the $k$-th direction. We represent both $\frac{\Delta}{\Delta x_k}$ and $\frac{\Delta^2}{\Delta x_k^2}$ using eight-order central finite difference stencils~\cite{Fornberg1988} in matrix product operator (MPO) form. For details on how the method of finite differences can be implemented in MPO form, see e.g.~\cite[p.~591]{Lubasch2018} or~\cite[p.~22]{Ripoll2021}. Further, note that the main text's variational field, $\mf{V}^*(\mf{r}_q) = \hat{\mf{e}}_1 u_1^*(\mf{r}_q) +\hat{\mf{e}}_2 u_2^*(\mf{r}_q)$, along with the previous field, $\mf{V}(\mf{r}_q) = \hat{\mf{e}}_1 u_1(\mf{r}_q) +\hat{\mf{e}}_2 u_2(\mf{r}_q)$, both lie within the MPS manifold $\mathcal{M}$ restricted at bond dimension $\chi$ (see Eq.~\eqref{eq:MPSvecFiel}).
For the sake of convenience, we employ linear algebra notation in the rest of this section. We introduce the vector $\mf{u}_i =\textnormal{vec}\left(u_i(\mf{r}_q)\right)$ populated by the values of $u_i$ at all grid points $\mf{r}_q$, and similarly for $\mf{u}^*_i = \textnormal{vec}\left(u_i^*(\mf{r}_q)\right)$, with 
$\mf{u}_i, \mf{u}^*_i \in \mathbb{R}^{2^{\Ns K} \times 1}$. Using this, we rewrite Eq.~(8) into
\begin{Sequation}\label{eq:costfunc0}
\begin{split}
\Theta(\mf{V}^*) =& \sum_{i,j=1}^2 \left\{ \mu \left( \frac{\Delta \mf{u}^*_i}{\Delta x_i} \right)^{t} \frac{\Delta \mf{u}^*_j}{\Delta x_j}\right\} + \sum_{i=1}^2 \Bigg\{\frac{(\mf{u}^*_i)^{t} \mf{u}^*_i}{\Delta t^2} + \frac{(\mf{u}^*_i)^{t}}{\Delta t}\left( \frac{- \mf{u}_i}{\Delta t} + \sum_{j=1}^2 \left\{ \mf{u}_j \frac{\Delta \mf{u}_i}{\Delta x_j} - \nu \frac{\Delta^2 \mf{u}_i}{\Delta x_j^2} \right\} \right) \\
+& \left( \frac{- \mf{u}_i}{\Delta t} + \sum_{j=1}^2 \left\{ \mf{u}_j \frac{\Delta \mf{u}_i}{\Delta x_j} - \nu \frac{\Delta^2 \mf{u}_i}{\Delta x_j^2} \right\} \right)^{t} \frac{\mf{u}^*_i}{\Delta t} \Bigg\} + \Bigg[... \Bigg],
\end{split}
\end{Sequation}
with $[...]$ representing the remaining part of $\Theta$ which is independent of $\mf{u}_i^*$ (it will vanish after differentiation), and $(\cdot)^t$ being the transpose. For the sake of simplicity, the nonlinear term is here represented in convective form, albeit it should be put into skew-symmetric form during actual numerical simulations~\cite{Zhang1991}. 

Let us now write out $\mf{V}^*$. Assuming $\mf{V}^*$ is put into mixed canonical form with canonical centre at site $n$, it can be written as
\begin{Sequation}
    \mf{V}^*(\mf{r}_q) = \sum_{i=1}^2 \hat{\mf{e}}_i \hat{A^*}^{\omega_1}_i \hat{A^*}^{\omega_2}_i\cdots \hat{A^*}^{\omega_{n-1}}_i C^{*\omega_{n}}_i \hat{B^*}^{\omega_{n+1}}_i \hat{B^*}^{\omega_{n+2}}_i \cdots \hat{B^*}^{\omega_{\Ns}}_i,
\end{Sequation}
with the left unitary $\hat{A^*}^{\omega_n}_i$ and right unitary $\hat{B^*}^{\omega_n}_i$ matrices satisfying orthonormality conditions like in Eq.~\eqref{canonical}. For ease of notation, we now define row and column vectors
\begin{Sequation}\begin{split}
 \Phi^{*\omega_1 \omega_2 \ldots \omega_{n-1}}_i &= \hat{A^*}^{\omega_1}_i \hat{A^*}^{\omega_2}_i \cdots \hat{A^*}^{\omega_{n-1}}_i \in \mathbb{R}^{1 \times d(n-1)}, \\
 \Psi^{*\omega_{n+1} \omega_{n+2} \ldots \omega_{\Ns}}_i &= \hat{B^*}^{\omega_{n+1}}_i \hat{B^*}^{\omega_{n+2}}_i \cdots \hat{B^*}^{\omega_{\Ns}}_i \in \mathbb{R}^{d(n) \times 1},
\end{split}\end{Sequation}
as well as matrices $\Phi_i^{*n} \in \mathbb{R}^{2^{K(n-1)} \times d(n-1)}$ and $\Psi_i^{*n} \in \mathbb{R}^{d(n) \times 2^{K(N-n)}}$,
obtained by stacking the vectors $\Phi^{*\omega_1 \omega_2 \ldots \omega_{n-1}}_i$ and $\Psi^{*\omega_{n+1} \omega_{n+2} \ldots \omega_{\Ns}}_i$ corresponding to all values of $\omega_1,\ldots,\omega_{\Ns}$.
Note that the unitarity of $\hat{A^*}^{\omega_n}_i$ and $\hat{B^*}^{\omega_n}_i$ leads also to $\Phi_i^{*n}$ and $\Psi_i^{*n}$ being unitary:
\begin{Sequation}\begin{split}\label{eq:uni}
  \left( \Phi_i^{*n} \right)^{t} \Phi_i^{*n} & = \mathds{1},  \\
  \Psi_i^{*n} \left( \Psi_i^{*n} \right)^{t}  & = \mathds{1}.
\end{split}\end{Sequation}
Moreover, let us define a vector $\mf{c}_i^{*n} = \textnormal{vec} \left(C_i^{*\omega_n}\right)$ of all values of $C_i^{*\omega_n}$.
Following a straightforward calculation, we can establish a linear map representation of the MPS:
\begin{Sequation}\label{eq:uni2}\begin{split}
\mf{u}_i^* = U_i^{*n} \mf{c}^{*n}_i, \quad \mbox{where} \quad U_i^{*n} = \Phi_i^{*n} \otimes \mathds{1} \otimes (\Psi_i^{*n})^t,
\end{split}\end{Sequation}
which holds for all $n$. Lastly, note that $\frac{\Delta}{\Delta x_i}$ is a skew-symmetric operator, $\left(\frac{\Delta}{\Delta x_i}\right)^t = - \frac{\Delta}{\Delta x_i}$. 

The above definitions allow us to rewrite Eq.~\eqref{eq:costfunc0} into
\begin{Sequation}
\begin{split}
\Theta(\mf{V}^*) =& \sum_{i,j=1}^2 \left\{ - \mu (\mf{c}^{*n}_i)^t \left(U_i^{*n} \right)^{t} \frac{\Delta}{\Delta x_i} \frac{\Delta}{\Delta x_j} U_i^{*n} \mf{c}_i^{*n} \right\}\\
&+\sum_{i=1}^2 \Bigg\{ \frac{(\mf{c}_i^{*n})^t \left(U_i^{*n} \right)^{t} U_i^{*n} \mf{c}_i^{*n} }{\Delta t} + \frac{(\mf{c}_i^{*n})^t\left(U_i^{*n}\right)^{t}}{\Delta t} \left( \frac{-\mf{u}_i}{\Delta t} + \sum_{j=1}^2 \left\{ \mf{u}_j \frac{\Delta \mf{u}_i}{\Delta x_j} - \nu \frac{\Delta^2 \mf{u}_i}{\Delta x_j^2}\right\} \right)\\
&\left( \frac{- \mf{u}_i}{\Delta t} + \sum_{j=1}^2 \left\{ \mf{u}_j \frac{\Delta \mf{u}_i}{\Delta x_j} - \nu \frac{\Delta^2 \mf{u}_i}{\Delta x_j^2}\right\} \right)^{t} \frac{ U_{i}^{*n} \mf{c}_i^{*n} }{\Delta t} \Bigg\}+ \Bigg[... \Bigg],
\end{split}
\end{Sequation}
with the canonical centre of $\mf{u}_i^*$ and its transpose set at site $n$. Now, the minimum of $\Theta(\mf{V}^*)$ is found at the stationary point where the gradient of $\Theta$ with regards to its variational variables vanishes. Here the variational variables in question are the matrices of $\mf{V}^*(\mf{r}_q)$. In other words, we require that
\begin{Sequation}\label{eq:dcostfunc0}
    \frac{\partial \Theta(\mf{V}^*)}{\partial C^{*\omega_n}_k} = 0
\end{Sequation}
simultaneously for all $n$ and $k$. The solution to Eq.~\eqref{eq:dcostfunc0} is given by
\begin{Sequation}\label{eq:dcostfunc1}
\begin{split}
    \left(U_i^{*n}\right)^{t} \Bigg( \sum_{j=1}^2 \left\{ -\mu \Delta t \frac{\Delta}{\Delta x_k} \frac{\Delta}{\Delta x_j} U_i^{*n} \mf{c}_i^{*n} \right\}+\frac{ U_i^{*n} \mf{c}_i^{*n} }{\Delta t} -\frac{\mf{u}_k}{\Delta t} + \sum_{j=1}^2 \left\{\mf{u}_j\frac{\Delta \mf{u}_k}{\Delta x_j} -\nu \frac{\Delta^2 \mf{u}_k}{\Delta x_j^2} \right\} \Bigg),
    \end{split}
\end{Sequation}
which can be rearranged into
\begin{Sequation}\label{eq:dcostfunc2}
\begin{split}
   \mf{c}_k^{*n} -\left(U_i^{*n} \right)^{t}\sum_{j=1}^2 \left\{ \mu \Delta t^2 \frac{\Delta}{\Delta x_k} \frac{\Delta}{\Delta x_j} U_i^{*n} \mf{c}_i^{*n} \right\} = \left(U_i^{*n} \right)^{t} \left( \mf{u}_k - \Delta t \sum_{j=1}^2 \left\{\mf{u}_j\frac{\Delta \mf{u}_k}{\Delta x_j} -\nu \frac{\Delta^2 \mf{u}_k}{\Delta x_j^2} \right\}\right),
   \end{split}
\end{Sequation}
after using Eqs.~\eqref{eq:uni} and~\eqref{eq:uni2} to set $\left(U_i^{*n} \right)^{t}  U_i^{*n} \mf{c}_i^{*n} = \mathds{1} \mf{c}_i^{*n} = \mf{c}_i^{*n}$.

\subsection{Local minimisation}\label{secMPS:B}
Solving Eq.~\eqref{eq:dcostfunc2} will locally minimise $\Theta(\mf{V}^*)$ with respect to the $n$-th tensor of the $k$-th flow component. We define
\begin{Sequation}\label{eq:redefines}
    \begin{split}
        &H_{kj} = \left(U_{i}^{*n} \right)^{t}\frac{\Delta}{\Delta x_k} \frac{\Delta}{\Delta x_j}U_i^{*n},\\
        &\boldsymbol\alpha_k = \mf{c}_k^{*n},\\
        &\boldsymbol\beta_k = \left(U_i^{*n} \right)^{t} \left(\mf{u}_k - \Delta t \sum_{j=1}^2 \left\{\mf{u}_j\frac{\Delta \mf{u}_k}{\Delta x_j} -\nu \frac{\Delta^2 \mf{u}_k}{\Delta x_j^2} \right\}\right),\\
    \end{split}
\end{Sequation}
and collect these into vectors $\boldsymbol\alpha=\textnormal{vec}(\boldsymbol\alpha_k),\boldsymbol\beta=\textnormal{vec}(\boldsymbol\beta_k)$ while constructing the matrix $H$ from elements $H_{kj}$. Inserting $\boldsymbol\alpha$, $\boldsymbol\beta$ and $H$ into Eq.~\eqref{eq:dcostfunc2} gives a set of linear equations for $\boldsymbol\alpha$
\begin{Sequation}\label{eq:linprob}
\left(\mathds{1} -  \mu \Delta t^2 H\right) \boldsymbol\alpha = \boldsymbol\beta,
\end{Sequation}
with $-H$ being a positive semi-definite matrix. This can be seen by considering
\begin{Sequation}
\boldsymbol\alpha^t H \boldsymbol\alpha = \sum_{i,j=1}^2 (\mf{u}_i^*)^t \frac{\Delta}{\Delta x_i} \frac{\Delta}{\Delta x_j}  \mf{u}_j^* = - \sum_{i,j=1}^2 \left( \frac{\Delta \mf{u}_i^*}{\Delta x_i}\right)^t \frac{\Delta \mf{u}_j^*}{\Delta x_j}.
\end{Sequation}
Because $\sum_{i,j=1}^2 \left( \frac{\Delta \mf{u}_i^*}{\Delta x_i}\right)^t \frac{\Delta \mf{u}_j^*}{\Delta x_j} = \left\| \overline{\nabla} \cdot \mf{V} \right\|^2_2 \geq 0$, $H$ must be negative semi-definite, making $\left(\mathds{1} -  \mu \Delta t^2 H\right)$ as a whole positive definite (recall that $\Delta t, \mu >0$). Furthermore, it can be shown that constructing $H$ costs $\mathcal{O}\left(\Ns \chi^4\right)$, while computing its action on $\boldsymbol\alpha$ costs just $\mathcal{O}\left(\Ns \chi^3\right)$. We therefore employed the iterative method of conjugate gradient descent (CGD)~\cite{Shewchuk1994} to solve the linear problem of Eq.~\eqref{eq:linprob}. 

\subsection{Global minimisation}\label{secMPS:C}
The global minimum of $\Theta$ can be obtained by adapting well-established  
techniques for calculating the ground state energy of quantum many-body systems. We 
start with $\mf{V}^*$ in right-canonical form and minimise $\Theta$ with respect  to 
matrix $n=1$ using Eq.~(\ref{eq:linprob}). Similarly to the density matrix 
renormalisation group (DMRG) algorithm~\cite{Schollwock2011}, we subsequently perform 
a QR decomposition on site $n=1$ and shift the canonical centre to site $n=2$, and 
optimise $\Theta$ with respect  to  $n=2$. We iterate this procedure and sweep through 
the sites of $\mf{V}^*$ until convergence is achieved.

\subsection{Theoretical computational scaling}\label{secMPS:D}
The computational complexity of our algorithm for each individual time-step is a product of the computational cost of the global sweeps multiplied with that of the local minimisations. Let \DIFdelbegin \DIFdel{$\langle m_1 \rangle$ }\DIFdelend \DIFaddbegin \DIFadd{$\overline{m}_1$ }\DIFaddend be the average number of sweeps required for the minimisation of $\Theta$ to converge, and $q$ be the cost associated with shifting the canonical centre as described in Sec.~\ref{secMPS:C}, and $c$ the computational cost of a local minimisation. Then the scaling will be \DIFdelbegin \DIFdel{$\mathcal{O}\left[\langle m_1 \rangle \Ns (q + c) \right]$ }\DIFdelend \DIFaddbegin \DIFadd{$\mathcal{O}\left[\overline{m}_1 \Ns (q + c) \right]$ }\DIFaddend because the cost of the sweeps scales with the number of matrices swept, which is $\Ns$. 

Shifting the canonical centre rightwards from matrix $n$ to $n+1$ (or leftwards, to $n-1$) is done in two parts. First, a QR decomposition is performed, followed by a tensor contraction between the upper triangular matrix and the next site. It is straightforward to show~\cite{Schollwock2011} that the cost of both goes as $\mathcal{O}(\chi^3)$, which in turn implies $q\sim \mathcal{O}(\chi^3)$.

The computational cost of the local minimisation equals the cost needed to first explicitly calculate $\boldsymbol\beta$ in Eq.~\eqref{eq:redefines} plus the subsequent cost associated with the CGD iterations needed to solve Eq.~\eqref{eq:linprob}. Regarding the latter, let \DIFdelbegin \DIFdel{$\langle m_2 \rangle$ }\DIFdelend \DIFaddbegin \DIFadd{$\overline{m}_2$ }\DIFaddend be the mean number of iterations CGD requires to converge to the solution of Eq. \eqref{eq:linprob} within the desired precision. The cost of these iterations is then on average a product of \DIFdelbegin \DIFdel{$\langle m_2 \rangle$ }\DIFdelend \DIFaddbegin \DIFadd{$\overline{m}_2$ }\DIFaddend and the cost of computing the action of $H$ on $\boldsymbol\alpha$. The last operation can be executed as a tensor contraction where the matrix $H$ is never explicitly formulated, but, instead, the tensor network of $H \boldsymbol\alpha$ is contracted using standard tensor contraction techniques~\cite{Orus2014} at $\mathcal{O}(\chi^3)$ cost. Calculating $\boldsymbol\beta$ is however more expensive due to each $\boldsymbol\beta_i$ containing the nonlinear $\sum_{j=1}^2 \mf{u}_j \frac{\Delta \mf{u}_i}{\Delta x_j}$. We (variationally) construct it using exact Hadamard products~\cite[p.~593]{Lubasch2018}, which involves tensor contractions scaling as $\mathcal{O}(\chi^4)$. This makes the cost of each local minimisation step go as \DIFdelbegin \DIFdel{$c \sim \chi^4 + \langle m_2 \rangle \chi^3$}\DIFdelend \DIFaddbegin \DIFadd{$c \sim \chi^4 + \overline{m}_2 \chi^3$}\DIFaddend . 

The above implies the computational complexity of each time-step to be 
\begin{Sequation}
    \mathcal{O}\left[\DIFdelbegin 
\DIFdel{m_1 }
\DIFdelend \DIFaddbegin \overline{m}\DIFadd{_1 }\DIFaddend \Ns (q + c) \right] = \mathcal{O}\left[ \DIFdelbegin 
\DIFdel{m_1 }
\DIFdelend \DIFaddbegin \overline{m}\DIFadd{_1 }\DIFaddend \Ns \left(\chi^4 + \chi^3(\textnormal{Const}+ \DIFdelbegin 
\DIFdel{m_2 }
\DIFdelend \DIFaddbegin \overline{m}\DIFadd{_2 }\DIFaddend \right) \right].
\end{Sequation}
The numerical precision is controlled by \DIFdelbegin \DIFdel{$\langle m_1\rangle$ and $\langle m_2 \rangle$}\DIFdelend \DIFaddbegin \DIFadd{$\overline{m}_1$ and $ \overline{m}_2$}\DIFaddend . However, in our experience, the number of sweeps or CGD iterations required to achieve a given precision does not change with increasing system size. Dropping these prefactors along with the non-dominant terms leads to the total computational complexity of
\begin{Sequation}\label{eq:compComplex1}
 \mathcal{O}\left[\DIFdelbegin 
\DIFdel{m_1 }
\DIFdel{\Ns (q }\DIFdelend \DIFaddbegin \overline{m}\DIFadd{_1 \Ns }\left(\DIFadd{\chi^4 }\DIFaddend + \DIFdelbegin \DIFdel{c) }\DIFdelend \DIFaddbegin \DIFadd{\chi^3(\textnormal{Const}+ }\overline{m}\DIFadd{_2}\right) \DIFaddend \right] = \mathcal{O}\left(\Ns \chi^4\right)
\end{Sequation}
per timestep. This scaling is equivalent to $\mathcal{O}(\chi^4 \log \Ng)$, as the number of grid points $\Ng$ is related with \DIFdelbegin \DIFdel{$\Ng$ }\DIFdelend \DIFaddbegin \DIFadd{$\Ns$ }\DIFaddend through $\Ng = 2^{K \Ns}$.


\subsection{Demonstration of computational scaling}\label{secMPS:E}
\begin{figure}
\centering
\includegraphics[width=1\linewidth]{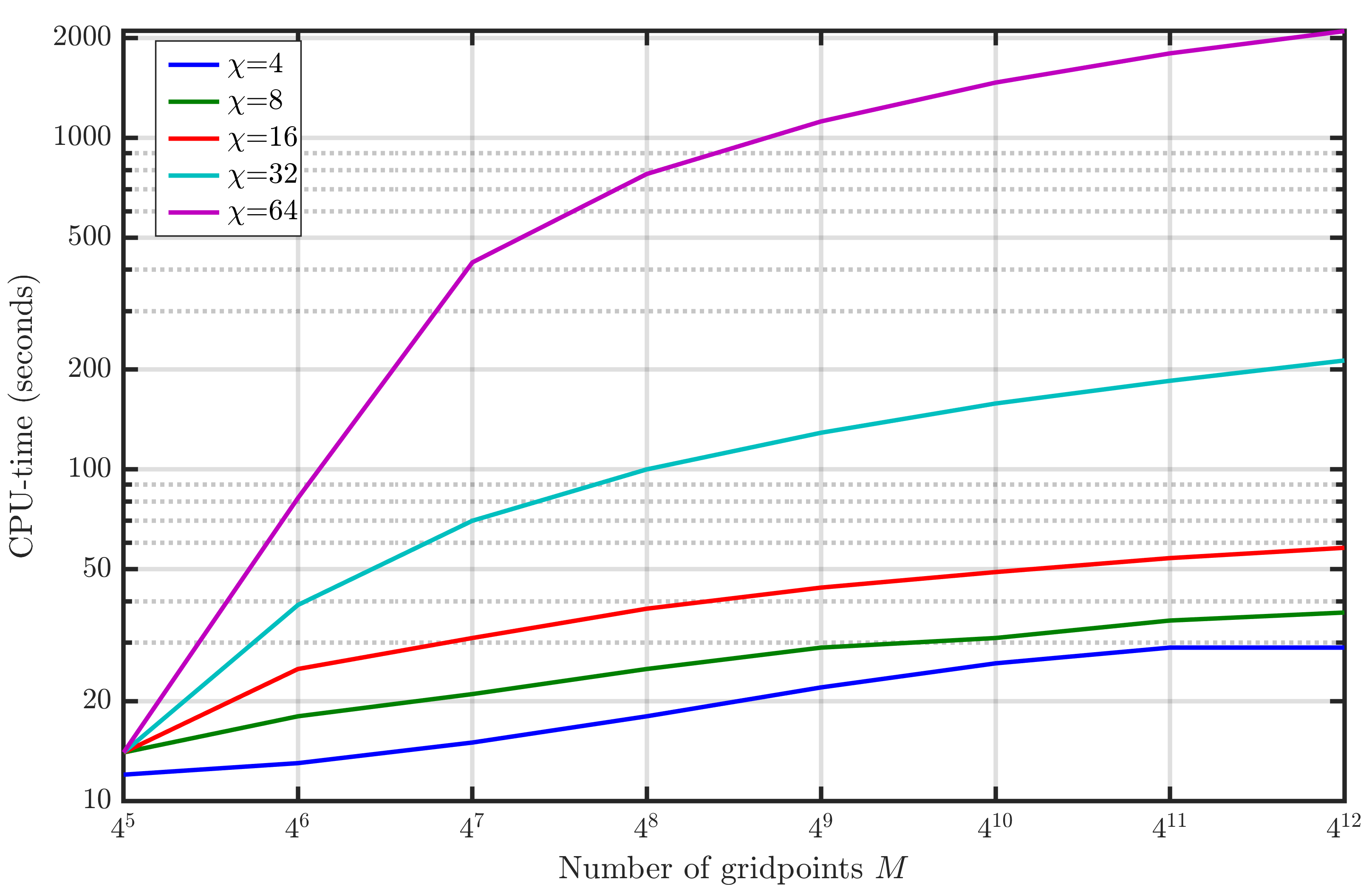}
\caption{\label{fig:perfDem}
\textbf{Demonstration of MPS computational scaling}. 10 time-steps of the 2-D Navier-Stokes equations were simulated with our MPS algorithm. The CPU time (in seconds) required to perform these 10 time-steps is plotted for five values of $\chi$ against eight different grid-sizes.
}
\end{figure}

Among the most interesting aspects of the MPS algorithm is how the computational complexity of the above Sec.~\ref{secMPS:D} scales linearly with $\Ns$, i.e., logarithmically with $\Ng$. To demonstrate, we have plotted in Supp. Fig.~\ref{fig:perfDem} the actual CPU time our algorithm requires to perform 10 time-steps simulating the 2-D Navier-Stokes equation at various values of $\Ng$ and $\chi$. The simulations were all carried out on a {\tt MacBook Pro (Retina, 15-inch, Mid 2015)} running {\tt macOS v. 11.6} on a {\tt 2,5 GHz Quad-Core i7 CPU} with {\tt 16 GB 1600 MHz DDR3 ram}. Note how all the curves in the figure saturate for sufficiently many gridpoints; this implies the computational performance scales \emph{sub-polynomially} with $\Ng$ in a manner consistent with the computational complexity derived in Eq.~\eqref{eq:compComplex1}.
\subsection{Arithmetic intensity}\label{MPS:F}
For completeness we also briefly discuss how memory-efficient our algorithm is. This can be done by studying a quantity known as the arithmetic intensity $I$. $I$ is the ratio between the amount of arithmetic operations performed (in units of FLOPs) to the required memory traffic (in bytes) of an algorithm. Many algorithms running on modern computing systems are memory bound in the sense that their performance is limited by memory bandwidth rather than computing power. Thus, algorithms with a high value of $I$ can more optimally utilize modern hardware than those with a low arithmetic intensity. 

As discussed in Sec.~\ref{secMPS:D}, our algorithm is dominated by the need to calculate the nonlinear term. This in practice boils down to repeatedly performing matrix-matrix multiplications between pairs of non-square matrices of type $(\textnormal{Const}\cdot\chi^2) \times \chi$ and $\chi \times (\textnormal{Const}\cdot\chi)$. Theoretically, the arithmetic intensity of such operations goes as
\begin{Sequation}\label{eq:ariInten}
I \sim \frac{\mathcal{O}(\chi^4)}{\mathcal{O}\left(\chi^2(\chi + \textnormal{Const})\right)} = \mathcal{O}(\chi)
\end{Sequation}
for large $\chi$ when memory is efficiently utilised. If, however, the memory is \emph{not} efficiently utilised due to shortcomings in either hardware (e.g. inadequate cache size) or software (e.g. failure to use cache blocking), the scaling in Eq.~\eqref{eq:ariInten} will not hold. 

Fortunately, the problem of matrix-matrix multiplication is among the most important and thoroughly studied problems of numerical linear algebra. Highly optimised hardware (along with the associated software) beyond CPUs exist for such operations, ranging from GPUs~\cite{Ernst2021} to even TPUs~\cite{Jouppi2018}. Thus, the dominant operation of our algorithm is characterised by a high $I$. This in turn makes our algorithm capable of efficiently utilising the power of modern parallelised computing hardware.

\clearpage
\newpage
\section{Quantum algorithm}\label{sec:quantumAlg}
Here we explain how to port our algorithm to a quantum computer. Our approach is based on the formalism of variational quantum algorithms (VQAs)~\cite{Cerezo2021} and summarized in Sec.~\ref{sec:VQA}. We compare this strategy to the alternative quantum algorithmic approaches~\cite{lloyd:20, Liu2021} in Sec.~\ref{sec:CompAlternQA}.

\subsection{Variational quantum algorithm}
\label{sec:VQA}

The main difference between the classical MPS algorithm presented in the main text and a corresponding VQA is that the latter encodes the solution in $2^{K\Ns}$ probability amplitudes of a quantum state $|u(\boldsymbol{\phi})\rangle$ of $K \Ns$ qubits. This state is created from a fixed initial state $|0\rangle^{\otimes K \Ns} = |\boldsymbol{0}\rangle$ by a network of quantum gates $\hat{U}(\boldsymbol{\phi})$ that are parameterised by classical variational parameters $\boldsymbol{\phi}$, i.e. $|u(\boldsymbol{\phi})\rangle = \hat{U}(\boldsymbol{\phi}) |\boldsymbol{0}\rangle$. Problem dependent gate operations and measurements are then applied to evaluate a cost function \cite{lubasch:20, Cerezo2021}.

\begin{figure}
\centering
\includegraphics[width=99.873mm]{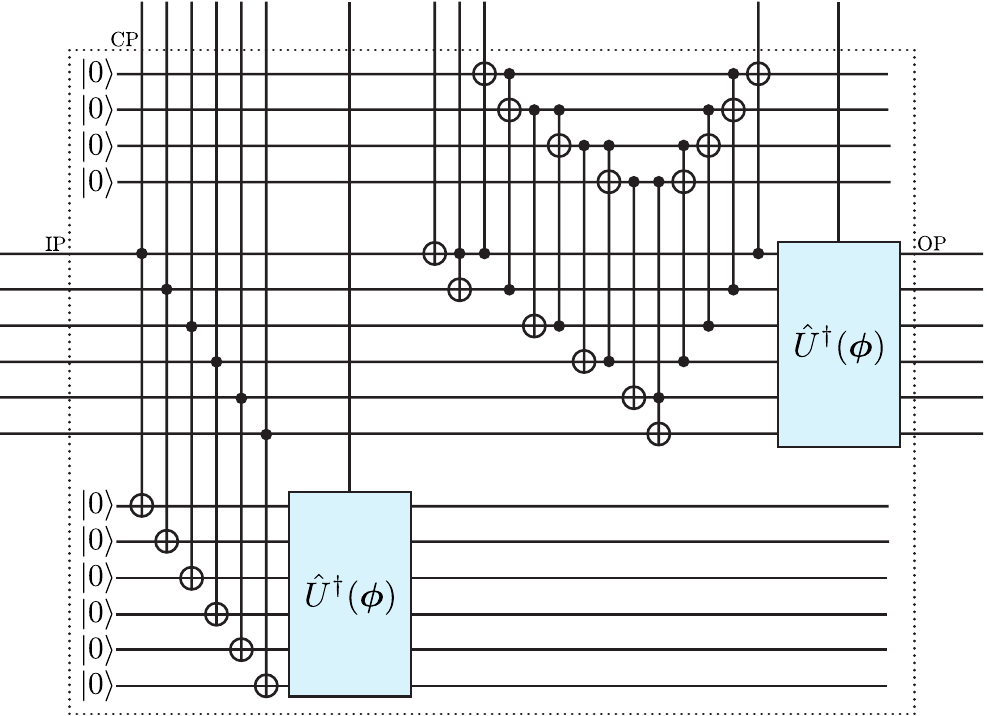}
\caption{\label{fig:VQC}
\textbf{Central part of the quantum circuit for the nonlinear term in Eq.~(\ref{eq:costGradientsVQA}).}
Gate $\hat{U}^{\dag}(\boldsymbol{\phi})$ is the adjoint of the quantum network creating the previous solution $|u\rangle$.
The variational state $|u^{*}\rangle$ is connected to the input port IP. The control port CP leads to the ancillary quantum control logic and the output port OP connects to the evaluation quantum logic discussed in detail in \cite{lubasch:20}.
Here the quantum state is realized with $\Ns = 6$ qubits.
}
\end{figure}

We sketch these problem dependent operations for the minimization of the cost function Eq.~\eqref{eq:costfunc0} for the special case of $K = 1$ and $\mu = 0$, i.e.\ Burgers' equation, as the more general Navier-Stokes case follows straightforwardly.
The quantum version of the cost function Eq.~\eqref{eq:costfunc0} is written as
\begin{Sequation}\begin{split}\label{eq:costVQA}
 \Theta(|u^{*}\rangle) & = \bigg\| \frac{|u^{*}\rangle - |u\rangle}{\Delta t} + \big(|u\rangle \cdot \overline{\nabla} \big) |u\rangle - \nu \overline{\nabla}^{2} |u\rangle \bigg\|_{2}^{2}.
\end{split}\end{Sequation}
Here $|u^{*}\rangle = |u(\boldsymbol{ \phi}^*)\rangle$ is the trial solution for the current time step and $|u\rangle = |u(\boldsymbol {\phi})\rangle$ is the solution from the previous time step. Efficient VQA methods for the minimization of Eq.~\eqref{eq:costVQA} use gradient-based optimizers~\cite{Cerezo2021} which require evaluation of only those scalar products in Eq.~\eqref{eq:costVQA} that contain $|u^{*}\rangle$. The gradient reads
\begin{Sequation}\begin{split}\label{eq:costGradientsVQA}
 \frac{\partial \Theta(|u^{*}\rangle)}{\partial \phi_{k}^{*}} & = \frac{\partial}{\partial \phi_{k}^{*}} \left( \frac{(\phi_{0}^{*})^{2}}{\Delta t^{2}} - \frac{2 \phi_{0} \phi_{0}^{*}}{\Delta t} \Re \left\{ \frac{1}{\Delta t} \langle u | u^{*} \rangle + \langle u | \big( |u\rangle \cdot \overline{\nabla} \big)^{\dag} |u^{*}\rangle - \nu \langle u | \overline{\nabla}^{2} |u^{*} \rangle \right\} \right)
\end{split}\end{Sequation}
where $\Re \{ \cdot \}$ is the real part, $(\cdot)^{\dag}$ the adjoint and $\phi_{0}$ and $\phi_{0}^{*}$ keep track of the changing normalization~\cite{lubasch:20}.
The quantum circuit representations of all terms in Eq.~\eqref{eq:costGradientsVQA} are given in~\cite[Supplementary Sec. III]{lubasch:20} together with a comprehensive derivation. Importantly, by making use of two copies of $|u\rangle$ we can straightforwardly handle the nonlinear term. The core part of the quantum network for evaluating the nonlinear term is shown in Supp. Fig.~\ref{fig:VQC}.

A distinguishing feature of the quantum version of our algorithm is that the variational states encoded in $|u(\boldsymbol{\phi})\rangle$ are not limited to MPS and can thus be more expressive. Therefore VQA optimization makes use of a variational manifold that is more general than the MPS manifold with the potential to lead to a quantum advantage \cite{lubasch:20}. Even when remaining within the MPS manifold the VQA approach immediately improves the scaling with bond dimension from $\mathcal{O}(\chi^4)$ to $\mathcal{O}(\chi^2)$~\cite{lubasch:20}.
\subsection{Comparison to alternative proposals}
\label{sec:CompAlternQA}

The VQA approach in Sec.~\ref{sec:VQA} and the alternative proposals~\cite{lloyd:20, Liu2021} have in common that they require a number of qubits that scales logarithmically with the number of grid points.
That is why all three approaches have the potential to achieve an exponential quantum speedup over some standard classical computational methods.

The main difference between the three strategies is that our approach concentrates on solving general nonlinear problems whereas the approaches~\cite{lloyd:20, Liu2021} focus on being efficient in the number of time steps.
The proposal~\cite{lloyd:20} is based on the derivation of the nonlinear Schr\"{o}dinger equation from the linear Schr\"{o}dinger equation for quantum many-body systems, i.e.\ a mean-field approximation, which is accurate only in the limit of weak interactions.
The approach~\cite{Liu2021} uses the well-known technique of Carleman linearization to map a specific set of weakly nonlinear ordinary differential equations with dissipation to a higher-dimensional linear problem.
Both methods~\cite{lloyd:20, Liu2021} solve their linear problems using the quantum linear systems algorithm~\cite{Harrow2009, Berry2014, Berry2017}.
This allows them to be efficient in the number of time steps and leads to mathematically rigorous convergence guarantees that, however, hold only under the strong restrictions of their derivation.
In contrast, our approach can be applied to a wide range of nonlinear partial differential equations with arbitrary interaction strengths.

\newpage \clearpage
\section{Graphical notation}\label{sec:graphical}
In this section we introduce graphical tensor network notation and use it to sketch MPS, tree tensor networks (TTNs) and the multiscale entanglement renormalisation ansatz (MERA). 

\begin{figure}[t!]\begin{center}
\includegraphics[width=0.5\linewidth]{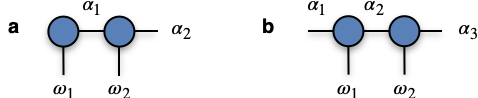} 
\caption{\label{fig:tenCon} \textbf{Tensor contractions}. Sub-figure ({\textbf a}) illustrates a tensor contraction between an order-2 tensor (left) with an order-3 tensor (right). The contraction is performed along the closed bond $\alpha_1$, while the open bonds $\omega_1, \omega_2$ and $\alpha_2$ are not summed over. A similar tensor contraction is shown in ({\textbf b}), except now both the tensors are order-3 and the summed-over bond is $\alpha_2$.}
\end{center}
\end{figure}

\begin{figure}[t!]\begin{center}
\includegraphics[width=0.6\linewidth]{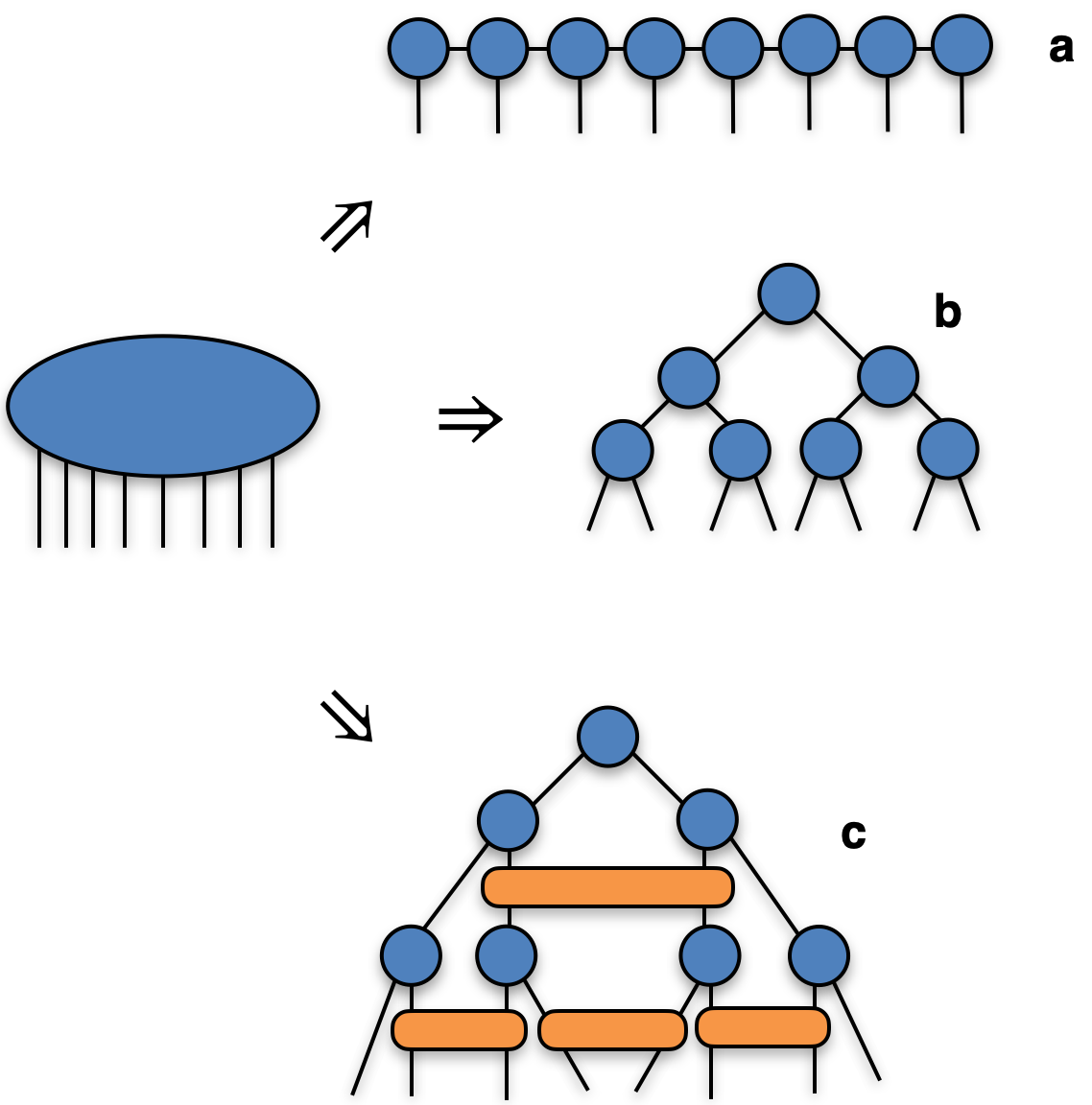} 
\caption{\label{fig:mpsttnmera} \textbf{Three possible decompositions from one tensor}. In ({\textbf a}), an order-8 tensor has been decomposed into an 8-site MPS where only the nearest neighbours are connected. In ({\textbf b}), the same tensor has been decomposed into a TTN, which connects distant sites. The MERA decomposition is shown in ({\textbf c}). Note both the increased connectivity compared to MPS and TTN, as well as the presence of loops.
}
\end{center}
\end{figure}

Describing tensor networks algebraically, as we did in Sec.~\ref{sec:mps} and beyond, is unpractical for geometries more complicated than MPS. Such tensor networks are more legibly described \emph{graphically}. To illustrate this graphical notation, consider the MPS in Eq.~\eqref{mpsF}. The matrix-matrix multiplications there can be written out as
\begin{Sequation}
v(\mf{r}_q,\chi) = \sum_{\{\alpha_n\}=1}^{d(n)} A^{\omega_1}_{\alpha_1}A^{\omega_2}_{\alpha_1 \alpha_2}\cdots A^{\omega_\Ns{-}1}_{\alpha_{N{-}2}\alpha_{N{-}1}} A^{\omega_\Ns}_{\alpha_{N{-}1}}.
\end{Sequation}
This equation emphasises that the matrix-matrix multiplication between $A^{\omega_{n}}$ and $A^{\omega_{n{+}1}}$ is equivalent to a summation over the internal index $\alpha_n$, which is equivalent to a tensor contraction. The tensor contraction can be graphically represented using the standard diagrammatic notation of tensor network theory (see e.g.~\cite{Schollwock2011,Orus2014}), as we do in Supp. Figs.~\ref{fig:tenCon}a and~\ref{fig:tenCon}b for $n = 1$ and $n = 2$, respectively. This notation allows us to \emph{draw} the MPS decomposition of Eq.~\ref{mpsF}, as shown for $N = 8$ in Supp. Fig.~\ref{fig:mpsttnmera}a.

Supp. Fig.~\ref{fig:mpsttnmera}a illustrates that the connectivity of MPS is solely between neighbouring sites. It is possible to connect more distant sites by using alternative tensor network geometries. One such geometry is the TTN~\cite{Shi2006,Tagliacozzo2009,Murg2010} that is illustrated in Supp. Fig.~\ref{fig:mpsttnmera}b. While TTNs connect distant sites, they also have fewer direct connections between neighbouring sites than MPS. The MERA network~\cite{Vidal2007} is a generalisation of TTNs that maintains connectivity between nearby sites, and is drawn in Supp. Fig~\ref{fig:mpsttnmera}c.

\newpage\clearpage

\bibliography{NikRefs}